\documentclass[reprint, superscriptaddress, amsmath, amssymb, aps, floatfix, nobibnotes,nofootinbib,longbibliography,prl]{revtex4-2}
\usepackage[latin9]{inputenc}
\usepackage{verbatim}
\usepackage{amsmath}
\usepackage{amssymb}
\usepackage{graphicx}

\usepackage{amsthm}
\usepackage{placeins}
\usepackage{booktabs}
\usepackage{array}
\usepackage{makecell}
\usepackage{epsfig}
\usepackage{subfigure}
\usepackage{float}
\usepackage{xcolor}
\usepackage{tikz}
\usepackage[normalem]{ulem}
\usepackage[colorlinks=true,linkcolor=blue,citecolor=blue]{hyperref}

\newcommand{\ket}[1]{|{#1})}
\newcommand{\bra}[1]{({#1}|}

\begin{document}

\title{
Exact superdiffusion exponents and hull exploration of chiral random walkers on critical percolation clusters
}

\author{Joshua Uhlig}
\affiliation{Institute of Physics and Astronomy, University of Potsdam,
14476 Potsdam, Germany}

\author{Ralf Metzler}
\email{ralf.metzler@uni-potsdam.de}
\affiliation{Institute of Physics and Astronomy, University of Potsdam,
14476 Potsdam, Germany}

\author{Erik Kalz}
\email{erikkalz98@gmail.com}
\affiliation{Institut de Science et d'Ing{\'e}nierie Supramol{\'e}culaires, University of Strasbourg, 67000 Strasbourg, France}

\begin{abstract}
Generalizing de Gennes' ant-in-the-labyrinth concept, we consider a chiral random walker on a percolation cluster at criticality. The fully chiral ant clings to obstacle boundaries and traces the fractal hulls of the cluster, spreading with the exact walk dimension $7/4$ until its orbit closes. Because the cluster contains holes on all scales, the ensemble spreads superdiffusively, $\langle r^2\rangle\sim t^{89/84}$, carried by a diminishing fraction of walkers whose orbits are still open. Both exponents follow from percolation geometry alone, and agree well with large-scale numerical simulations. When the walker turns chirally only with probability $\chi<1$, superdiffusion gives way to the subdiffusion of the ant at a crossover time set by both exponents, with a diffusion coefficient enhanced by $(1-\chi)^{-2/d_w}$. The relation between crossover time and diffusion coefficient does not depend on how chirality is realized, offering a direct test for chiral tracers in disordered media.
\end{abstract}

\maketitle

De Gennes' ant in the labyrinth~\cite{degennes1976percolation}, the random walk on a
percolation cluster, is the canonical idealization of transport in disordered media
\cite{havlin1987diffusion, benavraham2000diffusion} and is the basis for descriptions of
processes ranging from conduction in composites~\cite{kirkpatrick1973percolation} over
flow and transport in porous media and fractured rock~\cite{sahimi2011flow, berkowitz2002characterizing} to
the crowded interior of a living cell~\cite{saxton1994anomalous, hofling2013anomalous, sadegh2017plasma, mardoukhi2018fluctuations}. At the percolation threshold, the ant moves on the incipient infinite cluster (IIC), which is self-similar: its mass within a region of linear size $R$ grows as $R^{d_f}$, with fractal dimension $d_f=91/48$~\cite{dennijs1979relation, nienhuis1980magnetic, pearson1980conjecture, stauffer1992introduction}. On the IIC the ant is subdiffusive, $\langle r^{2}(t)\rangle \sim t^{2/d_{w}}$, with a walk dimension $d_{w} \approx 2.878$ that is known only numerically~\cite{alexander1982density, gefen1983anomalous, pandey1983confirmation, oshaughnessy1985analytical, grassberger1999conductivity}. The associated exponents define a robust universality class shared by all local, unbiased lattice walks on the IIC, while in continuum models the microdynamics can split it into several classes~\cite{spanner2016splitting}.

The random walking ant, per se, is memoryless: it
has no strategy for navigating the labyrinth and reacts to its surroundings purely at
random. The simplest strategy we can give the ant is the hand-on-the-wall rule~\cite{lucas1882recreations}, which solves any simply connected maze. It amounts to a persistent bias to turn in a fixed sense, that is, to chirality, a microscopic origin of so-called odd transport~\cite{hargus2021odd, fruchart2023odd,caprini2025active}. In fact, such a rotational bias occurs ubiquitously, from charged tracers in magnetic
fields and magnetic skyrmions to swimming bacteria and malaria parasites
threading disordered tissue
\cite{kalz2024oscillatory, reichhardt2022statics, lauga2006swimming, lettermann2025chirality,hargus2025odd}. In homogeneous media, chirality is known to suppress transport~\cite{wojcik2026chiral, hargus2021odd} and likewise heterogeneous environments hinder diffusion~\cite{rayleigh1892obstacles, havlin1987diffusion}, yet
near the percolation threshold, the two conspire to act against this intuition. Indeed, charges in a magnetic field and circle swimmers move along obstacle edges, enhancing transport and giving rise to localization transitions~\cite{schirmacher2015anomalous, siboni2018nonmonotonic, chepizhko2019ideal, chepizhko2020random, uhlig2026chirality}.

A minimal model for such motion is the chiral random walk (CRW)~\cite{wojcik2026chiral}, a lattice walker that at each step turns in a fixed sense with probability $\chi$ and otherwise moves in a random direction, interpolating between the ordinary ant ($\chi=0$) and a fully chiral walker, running in circles ($\chi=1$).
Above the percolation threshold the CRW has an optimal value of the chirality that maximizes the long-time diffusivity, set by a geometric matching between the persistence length of its wall-adhering motion and the typical obstacle cluster perimeter~\cite{uhlig2026chirality}. As the mean obstacle cluster size diverges, which by the matching-lattice duality of site percolation happens precisely at criticality~\cite{sykes1964exact}, the optimum shifts to the $\chi=1$ limit, i.e., to the wall-adhering motion~\cite{uhlig2026chirality}. Thus, the
fully chiral walker should be the optimal walker at critical percolation. Here we confirm this expectation and show that the chiral walker leaves the ant universality class altogether, with different exact exponents for individual trajectories and for the ensemble. 

Under the fully chiral rule ($\chi=1$) the
walker adheres to obstacle boundaries indefinitely and traces the
percolation hull, an object whose fractal dimension $d_h=7/4$ is known
exactly~\cite{saleur1987exact}. Each trajectory is therefore
superdiffusive, $r^2(t)\sim t^{2/d_h}=t^{8/7}$, with an exact walk
dimension $d_w^c=d_h=7/4$, as long as its orbit remains open. This places the fully chiral CRW in
the class of the hull-generating and kinetic walks constructed for this
purpose~\cite{ziff1984generation, majid1984kinetic, weinrib1985new, ziff1989hull} and of
Lorentz lattice gases with rotator-type scatterers~\cite{ziff1991lorentz}.
An ensemble of walkers, however, does not inherit this exponent, because orbits of different walkers around holes of all sizes close at all times.
We derive the exact exponent of the ensemble and study the case of chirality with randomness $(\chi<1)$ where at long times the ordinary random walk behavior is recovered. Chirality thus tunes the walk between two universality classes,
and we show that the percolation geometry fixes the exponents of the fully chiral
limit.

\newpage
\textit{Model}---We study the CRW, a discrete time and space random walk model for odd diffusion introduced in Ref.~\cite{wojcik2026chiral} on a 
two-dimensional square lattice of linear size $L$ with lattice constant and time
step both set to unity. 
The walker carries an internal state $d \in \{\rightarrow, \downarrow,
\leftarrow, \uparrow\}$ whose value corresponds to the direction of its most recent step. At each time step the
state is rotated to the next direction in a fixed cyclic order with
probability $\chi$, and redrawn uniformly from the four directions with probability $1-\chi$, after
which the walker moves in the indicated direction. 
Sites on the cluster are
independently accessible with occupation probability $p$ and blocked otherwise.
A walker stepping onto a blocked site is reflected to the site it previously occupied, which combined with the chiral rotation produces an
edge-adhering cycle: reflection reverses the direction, a first rotation
moves the walker one site along the boundary, and a second carries it
onto the next blocked site, so that the cycle repeats. We denote the walker's position at time $t$ by $\mathbf{r}(t)$ and characterize transport by the mean squared displacement (MSD) $\langle r^2(t)\rangle=\langle|\mathbf{r}(t)-\mathbf{r}(0)|^2\rangle$, averaged over walkers and disorder realizations, with positions unwrapped across the periodic boundaries.
Clusters are created at occupation probability $p=0.592746$, marginally below the site percolation threshold
$p_c=0.59274605079210(2)$~\cite{newman2000efficient, ziff2011results, jacobsen2015critical, mertens2022exact}, where the correlation length exceeds the system size by orders of magnitude and the largest cluster is indistinguishable from the IIC~\cite{havlin1987diffusion, stauffer1992introduction}. Walkers are initialized on sites of the
largest cluster, identified by the Hoshen--Kopelman
algorithm~\cite{hoshen1976percolation}. Fully chiral walkers are placed adjacent to at least one blocked site, since in open space a fully chiral walker circles a square of four neighboring sites indefinitely and thus never explores the cluster. Details of the model and simulations, including the effect of this initial condition, are given in the \hyperref[sec:endmatter]{End Matter}.

\begin{figure}
    \centering
    \includegraphics{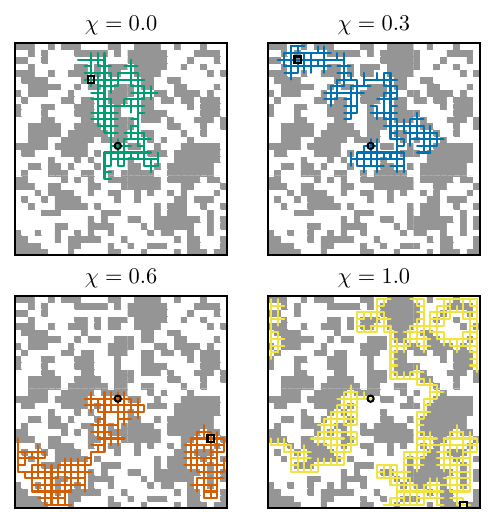}
    \caption{Trajectories of the chiral random walk on the same incipient infinite cluster at
$p = p_{c}$ ($L = 32$), for chiralities $\chi = 0$, $0.3$, $0.6$, and $1$. All
walkers start from the same site and are evolved for
$T = 10^{3}$ steps. Accessible sites are shown in white and obstacles in gray; black
circles mark the initial and black squares the final position. Spatial coverage grows
with $\chi$. At $\chi = 1$ (bottom right) the walker adheres to the boundary of a
single obstacle cluster and traces its hull.}
    \label{fig:traj}
\end{figure}

\textit{Hull tracing}---Figure~\ref{fig:traj} illustrates the effect of chirality. The ordinary achiral walker ($\chi=0$) explores a compact
neighborhood of its starting point, the hallmark of the recurrent, compact exploration
of a two-dimensional percolation cluster~\cite{condamin2007firstprl}. As $\chi$
increases, the trajectories cover a visibly larger portion of the cluster within the
same number of steps, and at $\chi = 1$ the walk collapses onto the boundary of a
single obstacle cluster, which it follows without ever detaching. This boundary is the hull of the obstacle cluster or,
equivalently, the boundary of the hole inside the accessible
cluster, and the fractal geometry of this boundary thus sets the walk dimension.

Since the walk adheres to the boundary, its step count is directly
convertible into an arc length along the hull. At $\chi=1$ every step
is a chiral rotation, so the edge-adhering cycle never breaks and the
walker advances along the boundary,
$\ell(t)\propto t$. For a self-similar curve of fractal dimension $d_h$, arc length and spatial extent are related by $\ell(R)\sim R^{d_h}$~\cite{feder1988fractals, falconer2003fractal}.
A walker on an unbounded hull therefore reaches a distance $R(t)\sim t^{1/d_h}$ from its starting point, so that its squared displacement obeys
\begin{equation}
  r^{2}(t)\sim t^{2/d_h}.
  \label{eq:exact_dw}
\end{equation}
With the exact hull dimension $d_h=7/4$~\cite{saleur1987exact}, this
yields $r^2(t)\sim t^{8/7}$ and a walk dimension $d_w^c=d_h=7/4$. These exponents are analytical, in contrast to the purely numerical value
$d_w\approx2.878$ of the achiral random walker. Chirality therefore changes the universality class of the ant in the labyrinth to one with exactly known exponents.

\begin{figure}
    \centering
\includegraphics{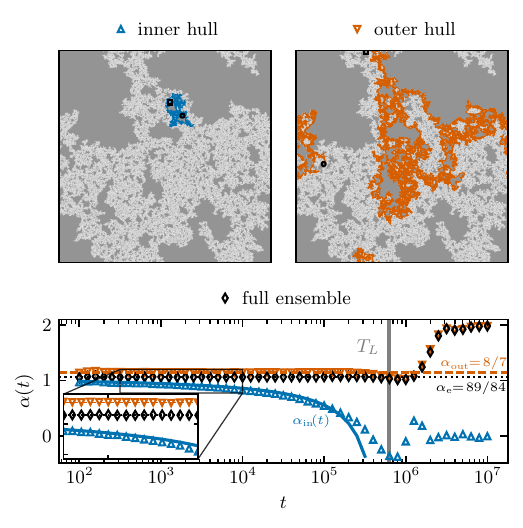}
    \caption{Fully chiral walkers ($\chi=1$) at $p=p_c$, $L=2048$. Top: trajectories over $T=10^6$ steps on an inner hull (left), whose
orbit closes, and on the outer hull (right), which winds around the
periodic system. Accessible sites white, obstacles gray; circles and
squares mark initial and final positions. Apparent variations in gray
are the unresolved structure of the cluster, visible on zooming in.
Bottom: local exponent $\alpha(t)=d\ln\langle r^2(t)\rangle/d\ln t$
over $T=10^7$ steps for outer hull, inner hull, and all walkers, with
$2/d_h=8/7$ (dashed red), $\alpha_{\mathrm{e}}=89/84$ (dotted black),
$\alpha_{\mathrm{in}}(t)$, Eq.~\eqref{eq:alpha_in} (solid blue), and
$T_L=L^{d_h}$ (gray vertical line).}
    \label{fig:hull}
\end{figure}

\textit{Ensemble exponent}---Figure~\ref{fig:hull} shows the local
exponent $\alpha(t)=d\ln\langle r^2(t)\rangle/d\ln t$ of the MSD for an ensemble of
fully chiral walkers started on the IIC. We show $t\geq100$, after an initial transient dominated by lattice discreteness. Walkers on the outer hull, the hull of the largest obstacle cluster, which typically wraps around the periodic system, follow Eq.~\eqref{eq:exact_dw} with $\alpha=2/d_h=8/7$ (red dashed line in Fig.~\ref{fig:hull}). The
full ensemble (black dotted line in Fig.~\ref{fig:hull}), however, shows a smaller value, and
walkers on inner hulls exhibit no plateau at all (blue solid line in Fig.~\ref{fig:hull}).

The reason for this difference is that Eq.~\eqref{eq:exact_dw} describes a walker on a
single, unbounded hull. On the IIC, the obstacle a walker starts next
to determines which hull it traces: next to a small obstacle cluster it
encircles a small hole, next to a large cluster a large hole, and since the
IIC contains holes of all sizes, the ensemble samples hulls on every
scale. A walker
on a hole of linear size $R_g$ spreads as $t^{2/d_h}$ only until it has
completed one orbit, after a time $T(R_g)\sim R_g^{d_h}$, and its squared
displacement then saturates at $\sim R_g^2$. We stress that this is not a finite-size
effect, as holes are finite on every scale even in an infinite system, due to the fractal nature. The
ensemble MSD therefore mixes walkers that are still spreading with
walkers whose orbits have closed, and its exponent is set by how the
walkers are distributed over hole sizes, a mixture that the time-averaged MSD of individual trajectories resolves directly (\hyperref[sec:endmatter]{End Matter}, Fig.~\ref{fig:tamsd}).

To obtain this distribution, we tile the cluster with boxes of linear
size $R$. Since the IIC has fractal dimension $d_f$, it occupies
$\sim(L/R)^{d_f}$ of them, each bordering $\mathcal{O}(1)$ holes of
size $\sim R$ with hull length $\sim R^{d_h}$, the IIC counterpart of the hull-length distribution of all perimeters~\cite{ziff1986test}. Holes of
this size thus contribute $\sim(L/R)^{d_f}R^{d_h}$ hull sites out of
the $\sim L^{d_f}$ sites of the cluster, such that a walker starting uniformly
on the cluster sits on such a hole with probability
$\sim R^{d_h-d_f}$. Summing over all scales larger than $R_g$ gives a geometric series that
converges because $d_h<d_f$, so that the fraction of walkers on holes
larger than $R_g$ scales like its first term,
\begin{equation}
  \Pi(R_g)\sim R_g^{\,d_h-d_f}=R_g^{-7/48}\,.
  \label{eq:Pi}
\end{equation}
Therefore,
$-(d\Pi/dR_g)\,dR_g$ is the fraction of walkers on holes of size between $R_g$ and
$R_g+dR_g$. At time $t$, walkers on holes larger than $R(t)\sim t^{1/d_h}$ are
still spreading with $r^2= R(t)^2$, while those on smaller holes
have saturated at $R_g^2$, so that
\begin{equation}
  \langle r^2(t)\rangle\sim
  \int_a^{R(t)}\!R_g^2\Big(-\frac{d\Pi}{dR_g}\Big)dR_g
  +R(t)^2\,\Pi\big(R(t)\big)
  \label{eq:msd_integral}
\end{equation}
and therefore $\langle r^2(t)\rangle\sim t^{\alpha_{\mathrm{e}}}$, with
\begin{equation}
  \alpha_{\mathrm{e}}=1+(2-d_f)/d_h=89/84>1,
  \label{eq:alphae}
\end{equation}
the exponent of the MSD of the full ensemble, in agreement with the plateau in
Fig.~\ref{fig:hull} (black dotted line). The exponent $\alpha_{\mathrm{e}}$ is
exact, yet distinct from the single-trajectory value
$2/d_h$. The growth of the MSD is effected by the vanishing fraction
$\Pi(R(t))\sim t^{(d_h-d_f)/d_h}= t^{-1/12}$ of walkers, those whose orbits are still open. The same counting predicts the number of distinct sites visited, $\langle S(t)\rangle\sim t\,\Pi\big(R(t)\big)\sim t^{1-(d_f-d_h)/d_h}=t^{11/12}$, which we confirm in the \hyperref[sec:endmatter]{End Matter}.

In a system of size $L$, the outer hull acts as a single hole of size $L$: it collects the walkers that would sit on larger holes in an infinite system. It carries the fraction
$\Pi(L)\sim L^{d_h-d_f}$ of the walkers, which keep spreading as
$t^{2/d_h}$ up to $T_L=L^{d_h}$, so the full ensemble is unaffected
by finite size effects before $T_L$. Among the walkers still spreading at time $t$,
however, the outer hull accounts for a fraction
\begin{equation}
  y(t)=\frac{\Pi(L)}{\Pi\big(R(t)\big)}
  =\left(\frac{t}{L^{d_h}}\right)^{(d_f-d_h)/d_h}
  =\left(\frac{t}{L^{d_h}}\right)^{1/12},
\end{equation}
which increases with time because the inner orbits keep closing while the outer one does not. Removing the outer hull therefore removes an increasing part of the
MSD, $\langle r^2(t)\rangle_{\mathrm{in}}\propto
t^{\alpha_{\mathrm{e}}}[1-y(t)]$, and the local exponent of the inner
hulls,
\begin{equation}
  \alpha_{\mathrm{in}}(t)=\alpha_{\mathrm{e}}
  -\Big(\frac{2}{d_h}-\alpha_{\mathrm{e}}\Big)\frac{y(t)}{1-y(t)},
  \label{eq:alpha_in}
\end{equation}
does not settle on a plateau but decreases continuously. Eq.~\eqref{eq:alpha_in} contains no free parameter and describes the inner hull data remarkably well (solid blue line in Fig.~\ref{fig:hull}) for $t\ll T_L$. For $t\gtrsim T_L$ (gray vertical line in Fig.~\ref{fig:hull}) the finite system takes over:
inner orbits have all closed ($\alpha_\mathrm{in}\to0$), and outer hull walkers wind around the
periodic system and drift ballistically ($\alpha_\mathrm{out}\to2$). The resulting finite-size
scaling forms are given in the \hyperref[sec:endmatter]{End Matter}.

\begin{figure}
    \centering
    \includegraphics{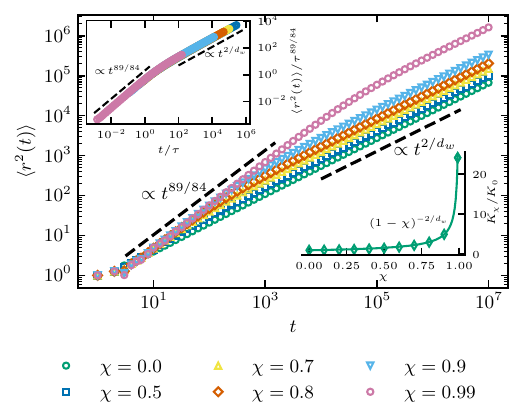}
    \caption{MSD at $p=p_c$ for chiralities $\chi\in[0.0,0.99]$.
Dashed lines indicate the superdiffusive $t^{\alpha_{\mathrm{e}}}$ and the subdiffusive $t^{2/d_w}$ scaling. Upper inset:
collapse with $\tau=(1-\chi)^{-2/(\alpha_{\mathrm{e}}d_w-2)}$,
(Eq.~\eqref{eq:tau}), displacement rescaled by
$\tau^{\alpha_{\mathrm{e}}}$. Lower inset: long-time amplitude
$K_\chi/K_0$ as a function of $\chi$; line:
$(1-\chi)^{-2/d_w}$ (Eq.~\eqref{eq:Kchi}), without free parameters.}
    \label{fig:crossover}
\end{figure}

\textit{Chirality with randomness}---For $\chi<1$ the direction of the walker is redrawn
with probability $1-\chi$ at every step, so the walker eventually
detaches from any hull with probability one.
Figure~\ref{fig:crossover} shows the MSD at $p_c$ for $\chi\in [0,0.99]$. Beyond a minimum chirality, the curves interpolate between
the two limits established above: at short times they exhibit the
ensemble exponent $\alpha_\mathrm{e}$ of the fully chiral walk
and at long times they cross over to the ordinary random walker scaling with $\alpha=2/d_w$. Unlike the
familiar crossover for regular walkers on a cluster above $p_c$, where a finite correlation length
separates an anomalous regime from ordinary diffusion, both regimes here
are anomalous and of opposite character: superdiffusion gives way to
subdiffusion, so that the walk changes dynamical universality class
partway through its own evolution. The crossover is determined by two features. Firstly, wherever the superdiffusive branch is visible, its
amplitude is independent of $\chi$,
$\langle r^2(t)\rangle= A\,t^{\alpha_{\mathrm{e}}}$. Secondly, the long-time branch recovers the random walk exponent, but it lies above the achiral curve by a factor that
grows with $\chi$ (lower inset of Fig.~\ref{fig:crossover}). To understand this enhancement, note that at long times the walker is
again an ordinary random walk on the IIC, with MSD
$\langle r^2\rangle|_\chi=K_\chi t^{2/d_w}$, where $K_\chi$ is the generalized diffusion coefficient. Between two randomizations, which occur on average every $n(\chi)=1/(1-\chi)$ steps, the walker moves deterministically along a hull. We propose that at long times the only effect of these runs is to accelerate the clock of the CRW by their mean length, $\langle r^2(t)\rangle|_\chi=\langle r^2(n(\chi)\,t)\rangle|_{\chi=0}$, and therefore

\begin{equation}
  K_\chi/K_0=n(\chi)^{2/d_w}=(1-\chi)^{-2/d_w}.
  \label{eq:Kchi}
\end{equation}
This expression contains no adjustable parameters, reduces to $K_0$ at $\chi=0$, and agrees very nicely with the measured amplitudes over the whole range of $\chi$ (green solid line in the lower right inset of Fig.~\ref{fig:crossover}).

Matching the $\chi$-independent short-time branch to Eq.~\eqref{eq:Kchi}
then fixes the crossover time, 
\begin{equation}
  \tau\propto(1-\chi)^{-2/(\alpha_{\mathrm{e}}d_w-2)}\approx(1-\chi)^{-1.91}\,,
  \label{eq:tau}
\end{equation}
which combines the exponents of both universality classes. The MSD
obeys $\langle r^2(t)\rangle=\tau^{\alpha_{\mathrm{e}}}
\mathcal{F}(t/\tau)$ with $\mathcal{F}(x)\sim x^{\alpha_{\mathrm{e}}}$
for $x\ll1$ and $x^{2/d_w}$ for $x\gg1$, as confirmed by the collapse in
the upper inset of Fig.~\ref{fig:crossover}. The matching also yields
$K_\chi\propto\tau^{\alpha_{\mathrm{e}}-2/d_w}$. Unlike Eq.~\eqref{eq:tau},
this relation does not rely on the assumption of  Eq.~\eqref{eq:Kchi}: it requires only the
$\chi$-independent short-time amplitude and the two exponents. Since it links two measurable quantities without involving $\chi$, it can be tested directly for chiral particles in disordered media. 

Equation~\eqref{eq:Kchi} is a hypothesis. A single run of $n$ steps along a fractal hull spans only $R\sim n^{1/d_h}$, and treating runs as independent jumps of that size would give the much weaker $K_\chi\propto(1-\chi)^{-(2/d_h-2/d_w)}$, which the data rule out. The data instead indicate that successive runs add up: because chirality fixes the sense of circulation, a walker that has taken a random step typically reattaches to the same hull and continues in the same direction, so that a detachment is more often a stutter than an escape into the bulk. The excellent agreement with the simulation results supports these assumptions.

\textit{Conclusion}---We studied the chiral random walk on the
critical percolation cluster and showed that a persistent turning bias
moves it out of the universality class of the ant in the labyrinth. At
$\chi=1$ the walker adheres indefinitely to obstacle boundaries, and
each trajectory traces a percolation hull with the exact walk dimension
$d_w^c=d_h=7/4$, instead of the numerically known $d_w\approx2.878$ of
the achiral ant. An ensemble of walkers does not inherit this exponent:
because holes are finite on all scales, orbits close at all times, and
the MSD grows with the distinct exponent
$\alpha_{\mathrm{e}}=1+(2-d_f)/d_h=89/84$, carried by the fraction of
walkers whose orbits are still open. Both exponents follow from fractal dimensions alone, and the same counting accounts for the continuously drifting exponent of walkers on inner hulls and for the number of distinct sites visited, $\langle S(t)\rangle\sim t^{11/12}$. A fully chiral walker thus learns about the cluster only through its hulls: its spectrum cannot be summarized by a spectral dimension but encodes only the distribution of hull lengths, so that, unlike the ordinary random walker~\cite{alexander1982density, kac1966can,lapidus1991can,lapidus1991fractal}, it ``hears'' the ``coastline'' of the fractal drum rather than its membrane.

For $\chi<1$ the walk returns from $t^{\alpha_{\mathrm{e}}}$ to the ant's $t^{2/d_w}$ after a crossover time $\tau\propto(1-\chi)^{-2/(\alpha_{\mathrm{e}}d_w-2)}$, and thus to the ant universality class, with an amplitude enhanced by $(1-\chi)^{-2/d_w}$. The amplitude follows from treating the walker as an ant whose clock is accelerated by its mean run length $1/(1-\chi)$ along a hull, a hypothesis that reproduces the generalized diffusion coefficient without adjustable parameters. The unbounded growth of the
enhancement as $\chi\to1$ contrasts with the optimal chirality at
intermediate $\chi$ found above threshold~\cite{uhlig2026chirality}, and
confirms the expectation that motivated this work: at criticality, where
hulls exist on all scales, the fully chiral walker is the optimal one. Chirality thus emerges as a control parameter for transport in disordered media, with an optimal value set by the medium's geometry, which chiral walkers achieve by maximizing both how far they spread and how many sites they visit.

The relevance of a persistent turning bias reaches well beyond the
present model. Chirality and percolation also pertain to living matter, where spinning microalgae assemble into a percolating chiral fluid~\cite{eisenmann2026chirality}. Moreover, percolation is used to quantify habitat connectivity in
fragmented landscapes~\cite{keitt1997detecting}, where boundary-following
movement succeeds or fails depending on the fragmentation
pattern~\cite{malanson1999ants}, in close analogy to our distinction
between inner and outer hulls. In theoretical
computer science, rotor-based walks are studied precisely because a
fixed local turning order changes how a walker covers a
graph~\cite{priezzhev1996eulerian, yanovski2003distributed}. In these
settings, however, the advantage is established qualitatively, by
simulation, or as a bound for arbitrary graphs. The critical percolation cluster is a rare case where it can be pinned
to exact exponents, providing analytical benchmarks for such settings.

Experimentally, the crossover from superdiffusion to subdiffusion offers a direct test of our results. Since
$K_\chi\propto\tau^{\alpha_{\mathrm{e}}-2/d_w}$ follows from matching the two regimes, it holds independently of
how $\chi$ maps onto a physical parameter and of the clock hypothesis, Eq.~\eqref{eq:Kchi}, so it can be checked for any
chiral tracer in a disordered medium close to threshold. The magnetic analogue was in fact realized in two-dimensional electron gases with lithographically patterned obstacle arrays, where the measured magnetoconductivity peaks once the cyclotron radius falls below the obstacle size, a feature that simulations trace back to electrons moving along obstacle contours~\cite{siboni2018nonmonotonic}. Colloidal model systems already realize random obstacle matrices with tunable density near the localization transition~\cite{skinner2013localization}, so that replacing the tracers by chiral ones would provide a direct test. Candidate systems range from chiral active colloids, granular chiral particles~\cite{chan2024chiral, caprini2025active}, and swimming microorganisms in obstacle
arrays~\cite{lauga2006swimming, lettermann2025chirality}, where bacteria in disordered porous media can already be followed at single-cell resolution~\cite{bhattacharjee2019bacterial}, to magnetic
skyrmions, whose gyrotropic coupling renders their thermal motion
odd-diffusive~\cite{schutte2014inertia, reichhardt2022statics,
kalz2024oscillatory}. For skyrmions, pinning defects are known to enhance rather than suppress
diffusion~\cite{rieger2023defect}, the same interplay of odd dynamics
and obstacles that underlies the enhancement found here and in Ref.~\cite{uhlig2026chirality}. Since pinning
landscapes can now be both measured and
written~\cite{brems2025realizing, kern2022deterministic}, a percolating
defect pattern could be designed directly.

\begin{acknowledgments}
\textit{Acknowledgments}---
  We thank Jan W{\'o}jcik, Narinder, and Nir Sherf for helpful discussions.
  RM acknowledges support from German Research Foundation (DFG, grant no. ME1535/22-1).
\end{acknowledgments}

\onecolumngrid
\section*{End Matter}

\label{sec:endmatter}
\twocolumngrid
\textit{Model}---The CRW models odd diffusion in discrete space and time (lattice constant $a$ and time step $\delta t$) by equipping the walker with an internal degree of freedom, whose state $d$
encodes the direction of the walker's most recent step. 
This state is updated by a mixture of two rules: with probability $\chi$ it is 
changed deterministically to the next direction in a fixed cyclic order ($\rightarrow,
\downarrow, \leftarrow, \uparrow, \rightarrow, \dots$), and with probability
$1-\chi$ it is redrawn uniformly at random. Next, the walker steps in the lattice
direction given by the updated internal state. Therefore, 
the CRW interpolates between a normal random walk ($\chi=0$) and a fully deterministic
circular trajectory ($\chi=1$).

Following Ref.~\cite{wojcik2026chiral} the state of the particle is written as a probability vector in the form
\begin{equation}
\ket{\rho(t)} = \sum_{\mathbf{r}, d} p_{\mathbf{r}, d}(t)\, \ket{\mathbf{r}} 
\otimes \ket{d},
\end{equation}
with rounded kets as notation for probability vectors, and $\otimes$ for
the tensor product combining the internal degree of freedom and spatial vector
spaces. Here, $\ket{\mathbf{r}} 
= \ket{x}\ket{y}\in\mathbb{Z}^2$ are the spatial basis vectors,
and $\ket{d}$ with $d \in\{\rightarrow, \leftarrow, \uparrow, \downarrow\}$, 
are the basis vectors of the internal degree of freedom.
The time evolution of the joint position-coin distribution is
\begin{equation}
    \ket{\rho(t+1)} = U\ket{\rho(t)} = S\,C\ket{\rho(t)},
    \quad
    C=\sum_{\mathbf{r}}\ket{\mathbf{r}}\bra{\mathbf{r}}\otimes C_{\mathbf{r}},
\end{equation}
where $S$ moves the walker according to its internal state, e.g.\ $S\ket{x,y,\uparrow}=\ket{x,y+1,\uparrow}$, and the coin $C$ acts on the internal state with a site-dependent operator $C_{\mathbf{r}}$. On accessible sites, $C_{\mathbf{r}}$ is the CRW coin
\begin{equation}
\label{eq:coin_crw}
    C_{\text{CRW}}=(1-\chi)\,C_{\text{rand}}+\chi\,C_{\text{chir}},
\end{equation}
where $\chi\in[0,1]$, the chirality, is the probability of the chiral rule. On blocked sites, $C_{\mathbf{r}}=C_{\text{refl}}$ reverses the internal state, so that a walker stepping onto an obstacle returns to its previous site on the next step~\cite{wojcik2026chiral}. Here $C_{\text{rand}}$ randomizes the internal state, $C_{\text{chir}}$ cyclically permutes it, and $C_{\text{refl}}$ reverses it,
\begin{equation}
  C_{\text{rand}}=\tfrac{1}{4}\Bigl(\begin{smallmatrix}
    1&1&1&1\\ 1&1&1&1\\ 1&1&1&1\\ 1&1&1&1
  \end{smallmatrix}\Bigr),\;
  C_{\text{chir}}=\Bigl(\begin{smallmatrix}
    0&0&1&0\\ 0&0&0&1\\ 0&1&0&0\\ 1&0&0&0
  \end{smallmatrix}\Bigr),\;
  C_{\text{refl}}=\Bigl(\begin{smallmatrix}
    0&1&0&0\\ 1&0&0&0\\ 0&0&0&1\\ 0&0&1&0
  \end{smallmatrix}\Bigr),
  \label{eq:coins}
\end{equation}
in the basis $(\rightarrow,\leftarrow,\uparrow,\downarrow)$. Without obstacles, the long-time diffusion tensor follows from a discrete Green--Kubo relation~\cite{hargus2021odd, wojcik2026chiral},
\begin{equation}
    \mathbf{D}_\text{CRW}=\frac{a^2}{4\delta t}\,\frac{1-\chi^2}{1+\chi^2}\begin{pmatrix}
        1& \frac{-2\chi}{1-\chi^2}\\[1ex]
        \frac{2\chi}{1-\chi^2}& 1
    \end{pmatrix},
\end{equation}
with antisymmetric off-diagonal elements, $D_{yx}=-D_{xy}$, the hallmark of odd diffusion~\cite{hargus2021odd, kalz2022collisions}.

\begin{figure}
    \centering
    \includegraphics{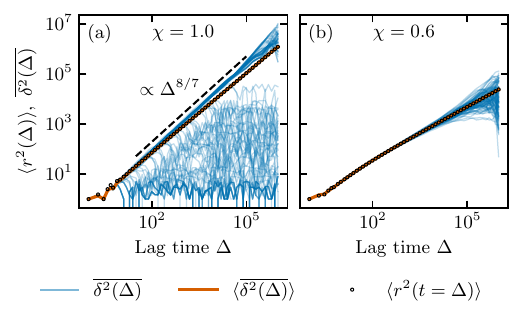}
    \caption{Time-averaged MSD, Eq.~\eqref{eq:tamsd}, at $p=p_c$,
$L=2048$, $T=10^6$. (a) Fully chiral walkers ($\chi=1$): individual TAMSDs
of $100$ randomly chosen trajectories (thin blue lines), their ensemble average
$\langle\overline{\delta^2(\Delta)}\rangle$ (red line) and the MSD (open circles).
Individual curves grow as $\Delta^{8/7}$ (dashed black line) and saturate
once their orbit has closed, with plateaus at all lag times.  (b) Same for $\chi=0.6$: individual TAMSDs coincide
with the MSD.}
    \label{fig:tamsd}
\end{figure}

\textit{Simulation Methods}---We simulate the chiral random walk on a square lattice of linear size $L$
with periodic boundary conditions, lattice constant $a = 1$, and time step
$\delta t = 1$. Each site is independently set to accessible with probability $p$ (site occupation probability)
and reflective otherwise. The largest connected cluster is identified using
the Hoshen--Kopelman algorithm~\cite{hoshen1976percolation}. For $\chi<1$ all walkers are
initialized on this cluster with random positions and uniformly distributed
initial internal degree of freedom states. In the fully chiral case $\chi=1$ we pick a random starting position on the largest cluster next to a blocked site and the initial internal degree of freedom to point away from the blocked site, as if a reflection just happened, so that edge-adhering can occur. 
This excludes only trivial trajectories: in open space a fully chiral walker circles a square of four neighboring sites indefinitely, with $r^2\le2a^2$. Starting uniformly on the whole cluster would add these walkers as a constant offset and reduce the amplitudes of $\langle r^2(t)\rangle$ and $\langle S(t)\rangle$ by the fraction of walkers that attach to a hull, leaving all exponents unchanged. Starting uniformly on hull-adjacent sites weights each hole by its hull length, as assumed in Eq.~\eqref{eq:Pi}.
For the outer-hull ensemble, walkers start on sites adjacent to the largest obstacle cluster. This cluster usually wraps around the periodic system; when it does not, its hull is a closed orbit of size $\sim L$, which behaves identically for $t\ll T_L$ and saturates rather than drifts afterwards. For the inner hull ensemble, on sites adjacent
to any other obstacle cluster that does not wrap around the periodic boundaries. Sites bordering both the outer and inner hull are excluded.
We simulate at $p = 0.592746$,
slightly below the critical threshold $p_c = 0.59274605079210(2)$~\cite{jacobsen2015critical, mertens2022exact},
at which the largest cluster is statistically indistinguishable from the incipient infinite cluster (IIC)~\cite{havlin1987diffusion,stauffer1992introduction}. For each disorder realization, we evolve $100$
independent trajectories for $T=10^7$ steps, repeated over $10^4$ realizations, yielding $10^6$
trajectories in total. With this choice, statistical errors are smaller than the symbol size in all figures and therefore not shown.

\begin{figure}
    \centering
    \includegraphics{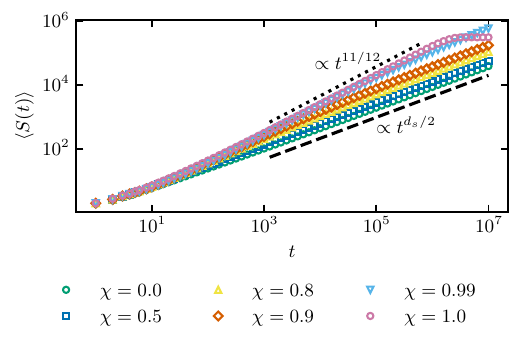}
    \caption{Mean number of distinct sites visited, $\langle S(t)\rangle$,
at $p=p_c$ ($L=2048$) for chiralities $\chi\in\{0.0,0.5,0.8,0.9,0.99,1.0\}$. Dashed line: $t^{d_s/2}=t^{d_f/d_w}\approx t^{0.659}$ of the ordinary
random walk, with spectral dimension $d_s=2d_f/d_w$; dotted line:
$t^{11/12}$ predicted for the fully chiral walker by the counting of
Eq.~\eqref{eq:Pi}, $\langle S(t)\rangle\sim t\,\Pi\big(R(t)\big)$.}
    \label{fig:S}
\end{figure}

\textit{Time-averaged MSD}---To resolve the individual trajectories
behind the ensemble exponent, we compute the time-averaged MSD (TAMSD) of
each trajectory~\cite{he2008random, metzler2014anomalous},
\begin{equation}
  \overline{\delta^2(\Delta)}=\frac{1}{T-\Delta}\sum_{t=0}^{T-\Delta-1}
  \big[\mathbf{r}(t+\Delta)-\mathbf{r}(t)\big]^2,
  \label{eq:tamsd}
\end{equation}
with lag time $\Delta$ and measurement time $T$. 
For $\chi<1$ the
individual TAMSDs coincide with each other and with the MSD for times $\Delta\ll T$ (Fig.~\ref{fig:tamsd} (b)). The randomizations detach the walker from any single hull, and the walk is ergodic, as for the normal random walker on the IIC~\cite{meroz2010subdiffusion, meroz2013test, mardoukhi2018fluctuations,cherstvy2018time}.
For $\chi=1$ the picture changes (Fig.~\ref{fig:tamsd} (a)). The
averaged TAMSD $\langle\overline{\delta^2(\Delta)}\rangle$ still coincides with
the MSD at all lag times. The individual TAMSDs,
however, each grow as
$\Delta^{2/d_h}=\Delta^{8/7}$ until their orbit closes at
$\Delta\sim R_g^{d_h}$ and then saturate at a plateau 
$\sim R_g^2$. Plateaus appear at all lag times, the direct signature
of holes on all scales, while walkers on the outer hull follow
$\Delta^{8/7}$. For the achiral random walk, comparable scatter near $p_c$ arises from walkers trapped on finite clusters, that is, from a disconnected phase space~\cite{mardoukhi2015geometry}. At $\chi=1$ the phase space, the IIC, is connected, and chirality partitions it into disjoint orbits. The scaling of the ensemble-averaged MSD is discussed in the main text.

\textit{Spectral dimension}---For the ordinary random walk, the spectral (fracton) dimension $d_s=2d_f/d_w$ links transport to the spectrum of the cluster: it governs the return probability $P_0(t)\sim t^{-d_s/2}$, the number of distinct sites visited $S(t)\sim t^{d_s/2}$, and the vibrational density of states $g(\omega)\sim\omega^{d_s-1}$~\cite{alexander1982density, rammal1983random, nakayama1994dynamical}.
At $\chi=1$ the evolution operator is a permutation of the walker's internal state and a walker returns to its starting state at multiples of its orbit period and after reflections, so that $P_0(t)$ is a comb of peaks at the orbit periods rather than a power law.
The spectrum therefore only represents the distribution of hull lengths.
This inverts the situation for drums with fractal rims, whose spectrum is set to leading order by the membrane and only corrected by the rim~\cite{lapidus1991can, lapidus1991fractal, sapoval1991vibrations}: the fully chiral walker ``hears'' the rim alone.
For walkers started on the IIC, the fraction on orbits longer than $\ell$ decays as $\Pi(\ell)\sim\ell^{-1/12}$, and a walker visits new sites at a constant rate until its orbit closes. Hence $\langle S(t)\rangle\sim t\,\Pi\big(R(t)\big)\sim t^{11/12}$, in agreement with the simulations, as shown in Fig.~\ref{fig:S}.
For $\chi<1$, $\langle S(t)\rangle$ crosses over from the fully chiral to the ordinary scaling.

\begin{figure}
    \centering
    \includegraphics{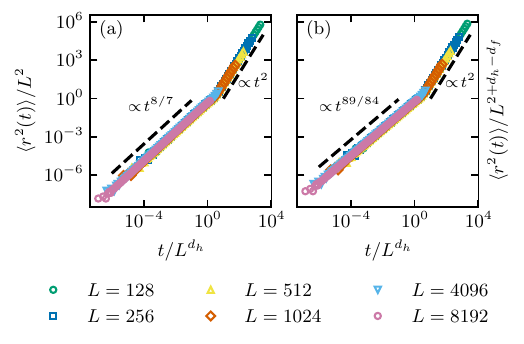}
    \caption{Finite-size scaling of the MSD for $\chi=1$ at $p=p_c$, for
$L\in\{128,256,512,1024,4096,8192\}$. (a) Outer hull walkers, rescaled as
$\langle r^2(t)\rangle/L^2$ against $t/L^{d_h}$, collapse onto
$\Phi(x)$ (Eq.~\eqref{eq:scaling_form}), with $\Phi\sim x^{8/7}$ before
and $\Phi\sim x^2$ after one circuit of the wrapping hull. (b) Full
ensemble, rescaled as $\langle r^2(t)\rangle/L^{2+d_h-d_f}$ against $t/L^{d_h}$ with
$2+d_h-d_f=89/48$, collapses onto $\Psi(x)$
(Eq.~\eqref{eq:scaling_full}), with $\Psi\sim x^{\alpha_{\mathrm{e}}}$
and $\Psi\sim x^2$.}
    \label{fig:collapse}
\end{figure}
\textit{Finite-size effects}---In a system of linear size $L$ the largest scale is the outer hull. When the largest obstacle cluster wraps around the periodic system, a walker on its hull winds around the system, and each circuit displaces the unwrapped trajectory by a lattice vector of length $L$. For $t\gg T_L$ the motion is therefore a deterministic drift,
  $\langle r^{2}(t)\rangle \sim L^{2}(t/L^{d_{h}})^{2}$
and together with Eq.~\eqref{eq:exact_dw} the outer hull MSD obeys
\begin{equation}
  \langle r^{2}(t)\rangle = L^{2}\,\Phi\!\left(\frac{t}{L^{d_h}}\right),
  \quad
  \Phi(x)\sim
  \begin{cases}
    x^{8/7}, & x\ll1,\\
    x^{2}, & x\gg1,
  \end{cases}
  \label{eq:scaling_form}
\end{equation}
with the crossover set by the hull length. Figure~\ref{fig:collapse}(a)
confirms this collapse for $L$ from $128$ to $8192$. Since the outer
hull carries a fraction $L^{d_h-d_f}$ of all walkers, the full ensemble obeys the analogous form
\begin{equation}
  \langle r^{2}(t)\rangle = L^{2+d_h-d_f}\,
  \Psi\!\left(\frac{t}{L^{d_h}}\right),
  \quad
  \Psi(x)\sim
  \begin{cases}
    x^{\alpha_{\mathrm{e}}}, & x\ll1,\\
    x^{2}, & x\gg1,
  \end{cases}
  \label{eq:scaling_full}
\end{equation}
with $2+d_h-d_f=89/48$ and Fig.~\ref{fig:collapse}(b) confirms this collapse.


\begin{thebibliography}{77}%
\makeatletter
\providecommand \@ifxundefined [1]{%
 \@ifx{#1\undefined}
}%
\providecommand \@ifnum [1]{%
 \ifnum #1\expandafter \@firstoftwo
 \else \expandafter \@secondoftwo
 \fi
}%
\providecommand \@ifx [1]{%
 \ifx #1\expandafter \@firstoftwo
 \else \expandafter \@secondoftwo
 \fi
}%
\providecommand \natexlab [1]{#1}%
\providecommand \enquote  [1]{``#1''}%
\providecommand \bibnamefont  [1]{#1}%
\providecommand \bibfnamefont [1]{#1}%
\providecommand \citenamefont [1]{#1}%
\providecommand \href@noop [0]{\@secondoftwo}%
\providecommand \href [0]{\begingroup \@sanitize@url \@href}%
\providecommand \@href[1]{\@@startlink{#1}\@@href}%
\providecommand \@@href[1]{\endgroup#1\@@endlink}%
\providecommand \@sanitize@url [0]{\catcode `\\12\catcode `\$12\catcode
  `\&12\catcode `\#12\catcode `\^12\catcode `\_12\catcode `\%12\relax}%
\providecommand \@@startlink[1]{}%
\providecommand \@@endlink[0]{}%
\providecommand \url  [0]{\begingroup\@sanitize@url \@url }%
\providecommand \@url [1]{\endgroup\@href {#1}{\urlprefix }}%
\providecommand \urlprefix  [0]{URL }%
\providecommand \Eprint [0]{\href }%
\providecommand \doibase [0]{https://doi.org/}%
\providecommand \selectlanguage [0]{\@gobble}%
\providecommand \bibinfo  [0]{\@secondoftwo}%
\providecommand \bibfield  [0]{\@secondoftwo}%
\providecommand \translation [1]{[#1]}%
\providecommand \BibitemOpen [0]{}%
\providecommand \bibitemStop [0]{}%
\providecommand \bibitemNoStop [0]{.\EOS\space}%
\providecommand \EOS [0]{\spacefactor3000\relax}%
\providecommand \BibitemShut  [1]{\csname bibitem#1\endcsname}%
\let\auto@bib@innerbib\@empty
%</preamble>
\bibitem [{\citenamefont {de~Gennes}(1976)}]{degennes1976percolation}%
  \BibitemOpen
  \bibfield  {author} {\bibinfo {author} {\bibfnamefont {P.-G.}\ \bibnamefont
  {de~Gennes}},\ }\bibfield  {title} {\bibinfo {title} {La percolation: un
  concept unificateur},\ }\href@noop {} {\bibfield  {journal} {\bibinfo
  {journal} {La Recherche}\ }\textbf {\bibinfo {volume} {7}},\ \bibinfo {pages}
  {919} (\bibinfo {year} {1976})}\BibitemShut {NoStop}%
\bibitem [{\citenamefont {Havlin}\ and\ \citenamefont
  {{ben-Avraham}}(1987)}]{havlin1987diffusion}%
  \BibitemOpen
  \bibfield  {author} {\bibinfo {author} {\bibfnamefont {S.}~\bibnamefont
  {Havlin}}\ and\ \bibinfo {author} {\bibfnamefont {D.}~\bibnamefont
  {{ben-Avraham}}},\ }\bibfield  {title} {\bibinfo {title} {Diffusion in
  disordered media},\ }\href {https://doi.org/10.1080/00018738700101072}
  {\bibfield  {journal} {\bibinfo  {journal} {Adv. Phys.}\ }\textbf {\bibinfo
  {volume} {36}},\ \bibinfo {pages} {695} (\bibinfo {year} {1987})}\BibitemShut
  {NoStop}%
\bibitem [{\citenamefont {{ben-Avraham}}\ and\ \citenamefont
  {Havlin}(2000)}]{benavraham2000diffusion}%
  \BibitemOpen
  \bibfield  {author} {\bibinfo {author} {\bibfnamefont {D.}~\bibnamefont
  {{ben-Avraham}}}\ and\ \bibinfo {author} {\bibfnamefont {S.}~\bibnamefont
  {Havlin}},\ }\href {https://doi.org/10.1017/CBO9780511605826} {\emph
  {\bibinfo {title} {Diffusion and Reactions in Fractals and Disordered
  Systems}}}\ (\bibinfo  {publisher} {Cambridge University Press},\ \bibinfo
  {address} {Cambridge, UK},\ \bibinfo {year} {2000})\BibitemShut {NoStop}%
\bibitem [{\citenamefont {Kirkpatrick}(1973)}]{kirkpatrick1973percolation}%
  \BibitemOpen
  \bibfield  {author} {\bibinfo {author} {\bibfnamefont {S.}~\bibnamefont
  {Kirkpatrick}},\ }\bibfield  {title} {\bibinfo {title} {Percolation and
  conduction},\ }\href {https://doi.org/10.1103/RevModPhys.45.574} {\bibfield
  {journal} {\bibinfo  {journal} {Rev. Mod. Phys.}\ }\textbf {\bibinfo {volume}
  {45}},\ \bibinfo {pages} {574} (\bibinfo {year} {1973})}\BibitemShut
  {NoStop}%
\bibitem [{\citenamefont {Sahimi}(2011)}]{sahimi2011flow}%
  \BibitemOpen
  \bibfield  {author} {\bibinfo {author} {\bibfnamefont {M.}~\bibnamefont
  {Sahimi}},\ }\href {https://doi.org/10.1002/9783527636693} {\emph {\bibinfo
  {title} {Flow and Transport in Porous Media and Fractured Rock}}}\ (\bibinfo
  {publisher} {Wiley-VCH Verlag \& Co},\ \bibinfo {address} {Weinheim, GER},\
  \bibinfo {year} {2011})\BibitemShut {NoStop}%
\bibitem [{\citenamefont {Berkowitz}(2002)}]{berkowitz2002characterizing}%
  \BibitemOpen
  \bibfield  {author} {\bibinfo {author} {\bibfnamefont {B.}~\bibnamefont
  {Berkowitz}},\ }\bibfield  {title} {\bibinfo {title} {Characterizing flow and
  transport in fractured geological media: A review},\ }\href
  {https://doi.org/https://doi.org/10.1016/S0309-1708(02)00042-8} {\bibfield
  {journal} {\bibinfo  {journal} {Adv. Water Resour.}\ }\textbf {\bibinfo
  {volume} {25}},\ \bibinfo {pages} {861} (\bibinfo {year} {2002})}\BibitemShut
  {NoStop}%
\bibitem [{\citenamefont {Saxton}(1996)}]{saxton1994anomalous}%
  \BibitemOpen
  \bibfield  {author} {\bibinfo {author} {\bibfnamefont {M.}~\bibnamefont
  {Saxton}},\ }\bibfield  {title} {\bibinfo {title} {Anomalous diffusion due to
  binding: a {M}onte {C}arlo study},\ }\href
  {https://doi.org/https://doi.org/10.1016/S0006-3495(96)79682-0} {\bibfield
  {journal} {\bibinfo  {journal} {Biophys. J.}\ }\textbf {\bibinfo {volume}
  {70}},\ \bibinfo {pages} {1250} (\bibinfo {year} {1996})}\BibitemShut
  {NoStop}%
\bibitem [{\citenamefont {H{\"o}fling}\ and\ \citenamefont
  {Franosch}(2013)}]{hofling2013anomalous}%
  \BibitemOpen
  \bibfield  {author} {\bibinfo {author} {\bibfnamefont {F.}~\bibnamefont
  {H{\"o}fling}}\ and\ \bibinfo {author} {\bibfnamefont {T.}~\bibnamefont
  {Franosch}},\ }\bibfield  {title} {\bibinfo {title} {Anomalous transport in
  the crowded world of biological cells},\ }\href
  {https://doi.org/10.1088/0034-4885/76/4/046602} {\bibfield  {journal}
  {\bibinfo  {journal} {Rep. Prog. Phys.}\ }\textbf {\bibinfo {volume} {76}},\
  \bibinfo {pages} {046602} (\bibinfo {year} {2013})}\BibitemShut {NoStop}%
\bibitem [{\citenamefont {Sadegh}\ \emph {et~al.}(2017)\citenamefont {Sadegh},
  \citenamefont {Higgins}, \citenamefont {Mannion}, \citenamefont {Tamkun},\
  and\ \citenamefont {Krapf}}]{sadegh2017plasma}%
  \BibitemOpen
  \bibfield  {author} {\bibinfo {author} {\bibfnamefont {S.}~\bibnamefont
  {Sadegh}}, \bibinfo {author} {\bibfnamefont {J.~L.}\ \bibnamefont {Higgins}},
  \bibinfo {author} {\bibfnamefont {P.~C.}\ \bibnamefont {Mannion}}, \bibinfo
  {author} {\bibfnamefont {M.~M.}\ \bibnamefont {Tamkun}},\ and\ \bibinfo
  {author} {\bibfnamefont {D.}~\bibnamefont {Krapf}},\ }\bibfield  {title}
  {\bibinfo {title} {Plasma membrane is compartmentalized by a self-similar
  cortical actin meshwork},\ }\href {https://doi.org/10.1103/PhysRevX.7.011031}
  {\bibfield  {journal} {\bibinfo  {journal} {Phys. Rev. X}\ }\textbf {\bibinfo
  {volume} {7}},\ \bibinfo {pages} {011031} (\bibinfo {year}
  {2017})}\BibitemShut {NoStop}%
\bibitem [{\citenamefont {Mardoukhi}\ \emph {et~al.}(2018)\citenamefont
  {Mardoukhi}, \citenamefont {Jeon}, \citenamefont {Chechkin},\ and\
  \citenamefont {Metzler}}]{mardoukhi2018fluctuations}%
  \BibitemOpen
  \bibfield  {author} {\bibinfo {author} {\bibfnamefont {Y.}~\bibnamefont
  {Mardoukhi}}, \bibinfo {author} {\bibfnamefont {J.-H.}\ \bibnamefont {Jeon}},
  \bibinfo {author} {\bibfnamefont {A.~V.}\ \bibnamefont {Chechkin}},\ and\
  \bibinfo {author} {\bibfnamefont {R.}~\bibnamefont {Metzler}},\ }\bibfield
  {title} {\bibinfo {title} {Fluctuations of random walks in critical random
  environments},\ }\href {https://doi.org/10.1039/c8cp03212b} {\bibfield
  {journal} {\bibinfo  {journal} {Phys. Chem. Chem. Phys.}\ }\textbf {\bibinfo
  {volume} {20}},\ \bibinfo {pages} {20427} (\bibinfo {year}
  {2018})}\BibitemShut {NoStop}%
\bibitem [{\citenamefont {den Nijs}(1979)}]{dennijs1979relation}%
  \BibitemOpen
  \bibfield  {author} {\bibinfo {author} {\bibfnamefont {M.~P.~M.}\
  \bibnamefont {den Nijs}},\ }\bibfield  {title} {\bibinfo {title} {A relation
  between the temperature exponents of the eight-vertex and q-state {P}otts
  model},\ }\href {https://doi.org/10.1088/0305-4470/12/10/030} {\bibfield
  {journal} {\bibinfo  {journal} {J. Phys. A}\ }\textbf {\bibinfo {volume}
  {12}},\ \bibinfo {pages} {1857} (\bibinfo {year} {1979})}\BibitemShut
  {NoStop}%
\bibitem [{\citenamefont {Nienhuis}\ \emph {et~al.}(1980)\citenamefont
  {Nienhuis}, \citenamefont {Riedel},\ and\ \citenamefont
  {Schick}}]{nienhuis1980magnetic}%
  \BibitemOpen
  \bibfield  {author} {\bibinfo {author} {\bibfnamefont {B.}~\bibnamefont
  {Nienhuis}}, \bibinfo {author} {\bibfnamefont {E.~K.}\ \bibnamefont
  {Riedel}},\ and\ \bibinfo {author} {\bibfnamefont {M.}~\bibnamefont
  {Schick}},\ }\bibfield  {title} {\bibinfo {title} {Magnetic exponents of the
  two-dimensional q-state {P}otts model},\ }\href
  {https://doi.org/10.1088/0305-4470/13/6/005} {\bibfield  {journal} {\bibinfo
  {journal} {J. Phys. A}\ }\textbf {\bibinfo {volume} {13}},\ \bibinfo {pages}
  {L189} (\bibinfo {year} {1980})}\BibitemShut {NoStop}%
\bibitem [{\citenamefont {Pearson}(1980)}]{pearson1980conjecture}%
  \BibitemOpen
  \bibfield  {author} {\bibinfo {author} {\bibfnamefont {R.~B.}\ \bibnamefont
  {Pearson}},\ }\bibfield  {title} {\bibinfo {title} {Conjecture for the
  extended {P}otts model magnetic eigenvalue},\ }\href
  {https://doi.org/10.1103/PhysRevB.22.2579} {\bibfield  {journal} {\bibinfo
  {journal} {Phys. Rev. B}\ }\textbf {\bibinfo {volume} {22}},\ \bibinfo
  {pages} {2579} (\bibinfo {year} {1980})}\BibitemShut {NoStop}%
\bibitem [{\citenamefont {Stauffer}\ and\ \citenamefont
  {Aharony}(1992)}]{stauffer1992introduction}%
  \BibitemOpen
  \bibfield  {author} {\bibinfo {author} {\bibfnamefont {D.}~\bibnamefont
  {Stauffer}}\ and\ \bibinfo {author} {\bibfnamefont {A.}~\bibnamefont
  {Aharony}},\ }\href {https://doi.org/10.1201/9781315274386} {\emph {\bibinfo
  {title} {Introduction to percolation theory}}}\ (\bibinfo  {publisher}
  {Taylor \& Francis},\ \bibinfo {address} {London},\ \bibinfo {year}
  {1992})\BibitemShut {NoStop}%
\bibitem [{\citenamefont {Alexander}\ and\ \citenamefont
  {Orbach}(1982)}]{alexander1982density}%
  \BibitemOpen
  \bibfield  {author} {\bibinfo {author} {\bibfnamefont {S.}~\bibnamefont
  {Alexander}}\ and\ \bibinfo {author} {\bibfnamefont {R.}~\bibnamefont
  {Orbach}},\ }\bibfield  {title} {\bibinfo {title} {Density of states on
  fractals: {``Fractons''}},\ }\href
  {https://doi.org/10.1051/jphyslet:019820043017062500} {\bibfield  {journal}
  {\bibinfo  {journal} {J. Physique Lett.}\ }\textbf {\bibinfo {volume} {43}},\
  \bibinfo {pages} {625} (\bibinfo {year} {1982})}\BibitemShut {NoStop}%
\bibitem [{\citenamefont {Gefen}\ \emph {et~al.}(1983)\citenamefont {Gefen},
  \citenamefont {Aharony},\ and\ \citenamefont
  {Alexander}}]{gefen1983anomalous}%
  \BibitemOpen
  \bibfield  {author} {\bibinfo {author} {\bibfnamefont {Y.}~\bibnamefont
  {Gefen}}, \bibinfo {author} {\bibfnamefont {A.}~\bibnamefont {Aharony}},\
  and\ \bibinfo {author} {\bibfnamefont {S.}~\bibnamefont {Alexander}},\
  }\bibfield  {title} {\bibinfo {title} {Anomalous diffusion on percolating
  clusters},\ }\href {https://doi.org/10.1103/PhysRevLett.50.77} {\bibfield
  {journal} {\bibinfo  {journal} {Phys. Rev. Lett.}\ }\textbf {\bibinfo
  {volume} {50}},\ \bibinfo {pages} {77} (\bibinfo {year} {1983})}\BibitemShut
  {NoStop}%
\bibitem [{\citenamefont {Pandey}\ and\ \citenamefont
  {Stauffer}(1983)}]{pandey1983confirmation}%
  \BibitemOpen
  \bibfield  {author} {\bibinfo {author} {\bibfnamefont {R.~B.}\ \bibnamefont
  {Pandey}}\ and\ \bibinfo {author} {\bibfnamefont {D.}~\bibnamefont
  {Stauffer}},\ }\bibfield  {title} {\bibinfo {title} {Confirmation of
  dynamical scaling at the percolation threshold},\ }\href
  {https://doi.org/10.1103/PhysRevLett.51.527} {\bibfield  {journal} {\bibinfo
  {journal} {Phys. Rev. Lett.}\ }\textbf {\bibinfo {volume} {51}},\ \bibinfo
  {pages} {527} (\bibinfo {year} {1983})}\BibitemShut {NoStop}%
\bibitem [{\citenamefont {O'Shaughnessy}\ and\ \citenamefont
  {Procaccia}(1985)}]{oshaughnessy1985analytical}%
  \BibitemOpen
  \bibfield  {author} {\bibinfo {author} {\bibfnamefont {B.}~\bibnamefont
  {O'Shaughnessy}}\ and\ \bibinfo {author} {\bibfnamefont {I.}~\bibnamefont
  {Procaccia}},\ }\bibfield  {title} {\bibinfo {title} {Analytical solutions
  for diffusion on fractal objects},\ }\href
  {https://doi.org/10.1103/PhysRevLett.54.455} {\bibfield  {journal} {\bibinfo
  {journal} {Phys. Rev. Lett.}\ }\textbf {\bibinfo {volume} {54}},\ \bibinfo
  {pages} {455} (\bibinfo {year} {1985})}\BibitemShut {NoStop}%
\bibitem [{\citenamefont {Grassberger}(1999)}]{grassberger1999conductivity}%
  \BibitemOpen
  \bibfield  {author} {\bibinfo {author} {\bibfnamefont {P.}~\bibnamefont
  {Grassberger}},\ }\bibfield  {title} {\bibinfo {title} {Conductivity exponent
  and backbone dimension in 2-d percolation},\ }\href
  {https://doi.org/10.1016/S0378-4371(98)00435-X} {\bibfield  {journal}
  {\bibinfo  {journal} {Physica A}\ }\textbf {\bibinfo {volume} {262}},\
  \bibinfo {pages} {251} (\bibinfo {year} {1999})}\BibitemShut {NoStop}%
\bibitem [{\citenamefont {Spanner}\ \emph {et~al.}(2016)\citenamefont
  {Spanner}, \citenamefont {H{\"o}fling}, \citenamefont {Kapfer}, \citenamefont
  {Mecke}, \citenamefont {Schr{\"o}der-Turk},\ and\ \citenamefont
  {Franosch}}]{spanner2016splitting}%
  \BibitemOpen
  \bibfield  {author} {\bibinfo {author} {\bibfnamefont {M.}~\bibnamefont
  {Spanner}}, \bibinfo {author} {\bibfnamefont {F.}~\bibnamefont
  {H{\"o}fling}}, \bibinfo {author} {\bibfnamefont {S.~C.}\ \bibnamefont
  {Kapfer}}, \bibinfo {author} {\bibfnamefont {K.~R.}\ \bibnamefont {Mecke}},
  \bibinfo {author} {\bibfnamefont {G.~E.}\ \bibnamefont {Schr{\"o}der-Turk}},\
  and\ \bibinfo {author} {\bibfnamefont {T.}~\bibnamefont {Franosch}},\
  }\bibfield  {title} {\bibinfo {title} {Splitting of the universality class of
  anomalous transport in crowded media},\ }\href
  {https://doi.org/10.1103/PhysRevLett.116.060601} {\bibfield  {journal}
  {\bibinfo  {journal} {Phys. Rev. Lett.}\ }\textbf {\bibinfo {volume} {116}},\
  \bibinfo {pages} {060601} (\bibinfo {year} {2016})}\BibitemShut {NoStop}%
\bibitem [{\citenamefont {Lucas}(1882)}]{lucas1882recreations}%
  \BibitemOpen
  \bibfield  {author} {\bibinfo {author} {\bibfnamefont {{\'E}.}~\bibnamefont
  {Lucas}},\ }\href@noop {} {\emph {\bibinfo {title} {{R{\'e}cr{\'e}ations
  math{\'e}matiques}}}},\ Vol.~\bibinfo {volume} {1}\ (\bibinfo  {publisher}
  {Gauthier-Villars},\ \bibinfo {address} {Paris},\ \bibinfo {year}
  {1882})\BibitemShut {NoStop}%
\bibitem [{\citenamefont {Hargus}\ \emph {et~al.}(2021)\citenamefont {Hargus},
  \citenamefont {Epstein},\ and\ \citenamefont {Mandadapu}}]{hargus2021odd}%
  \BibitemOpen
  \bibfield  {author} {\bibinfo {author} {\bibfnamefont {C.}~\bibnamefont
  {Hargus}}, \bibinfo {author} {\bibfnamefont {J.~M.}\ \bibnamefont
  {Epstein}},\ and\ \bibinfo {author} {\bibfnamefont {K.~K.}\ \bibnamefont
  {Mandadapu}},\ }\bibfield  {title} {\bibinfo {title} {Odd diffusivity of
  chiral random motion},\ }\href
  {https://doi.org/10.1103/PhysRevLett.127.178001} {\bibfield  {journal}
  {\bibinfo  {journal} {Phys. Rev. Lett.}\ }\textbf {\bibinfo {volume} {127}},\
  \bibinfo {pages} {178001} (\bibinfo {year} {2021})}\BibitemShut {NoStop}%
\bibitem [{\citenamefont {Fruchart}\ \emph {et~al.}(2023)\citenamefont
  {Fruchart}, \citenamefont {Scheibner},\ and\ \citenamefont
  {Vitelli}}]{fruchart2023odd}%
  \BibitemOpen
  \bibfield  {author} {\bibinfo {author} {\bibfnamefont {M.}~\bibnamefont
  {Fruchart}}, \bibinfo {author} {\bibfnamefont {C.}~\bibnamefont
  {Scheibner}},\ and\ \bibinfo {author} {\bibfnamefont {V.}~\bibnamefont
  {Vitelli}},\ }\bibfield  {title} {\bibinfo {title} {Odd viscosity and odd
  elasticity},\ }\href
  {https://doi.org/10.1146/annurev-conmatphys-040821-125506} {\bibfield
  {journal} {\bibinfo  {journal} {Annu. Rev. Condens. Matter Phys.}\ }\textbf
  {\bibinfo {volume} {14}},\ \bibinfo {pages} {471} (\bibinfo {year}
  {2023})}\BibitemShut {NoStop}%
\bibitem [{\citenamefont {Caprini}\ \emph {et~al.}(2026)\citenamefont
  {Caprini}, \citenamefont {Musacchio}, \citenamefont {Marconi}, \citenamefont
  {Liebchen},\ and\ \citenamefont {L{\"o}wen}}]{caprini2025active}%
  \BibitemOpen
  \bibfield  {author} {\bibinfo {author} {\bibfnamefont {L.}~\bibnamefont
  {Caprini}}, \bibinfo {author} {\bibfnamefont {M.}~\bibnamefont {Musacchio}},
  \bibinfo {author} {\bibfnamefont {U.~M.~B.}\ \bibnamefont {Marconi}},
  \bibinfo {author} {\bibfnamefont {B.}~\bibnamefont {Liebchen}},\ and\
  \bibinfo {author} {\bibfnamefont {H.}~\bibnamefont {L{\"o}wen}},\ }\bibfield
  {title} {\bibinfo {title} {Active thermodynamics of inertial chiral active
  gases: Equation of state and edge currents},\ }\href
  {https://doi.org/10.1073/pnas.2532158123} {\bibfield  {journal} {\bibinfo
  {journal} {Proc. Natl. Acad. Sci. USA}\ }\textbf {\bibinfo {volume} {123}},\
  \bibinfo {pages} {e2532158123} (\bibinfo {year} {2026})}\BibitemShut
  {NoStop}%
\bibitem [{\citenamefont {Kalz}\ \emph {et~al.}(2024)\citenamefont {Kalz},
  \citenamefont {Vuijk}, \citenamefont {Sommer}, \citenamefont {Metzler},\ and\
  \citenamefont {Sharma}}]{kalz2024oscillatory}%
  \BibitemOpen
  \bibfield  {author} {\bibinfo {author} {\bibfnamefont {E.}~\bibnamefont
  {Kalz}}, \bibinfo {author} {\bibfnamefont {H.~D.}\ \bibnamefont {Vuijk}},
  \bibinfo {author} {\bibfnamefont {J.-U.}\ \bibnamefont {Sommer}}, \bibinfo
  {author} {\bibfnamefont {R.}~\bibnamefont {Metzler}},\ and\ \bibinfo {author}
  {\bibfnamefont {A.}~\bibnamefont {Sharma}},\ }\bibfield  {title} {\bibinfo
  {title} {Oscillatory force autocorrelations in equilibrium odd-diffusive
  systems},\ }\href {https://doi.org/10.1103/PhysRevLett.132.057102} {\bibfield
   {journal} {\bibinfo  {journal} {Phys. Rev. Lett.}\ }\textbf {\bibinfo
  {volume} {132}},\ \bibinfo {pages} {057102} (\bibinfo {year}
  {2024})}\BibitemShut {NoStop}%
\bibitem [{\citenamefont {Reichhardt}\ \emph {et~al.}(2022)\citenamefont
  {Reichhardt}, \citenamefont {Reichhardt},\ and\ \citenamefont
  {Milo{\v{s}}evi{\'c}}}]{reichhardt2022statics}%
  \BibitemOpen
  \bibfield  {author} {\bibinfo {author} {\bibfnamefont {C.}~\bibnamefont
  {Reichhardt}}, \bibinfo {author} {\bibfnamefont {C.~J.~O.}\ \bibnamefont
  {Reichhardt}},\ and\ \bibinfo {author} {\bibfnamefont {M.~V.}\ \bibnamefont
  {Milo{\v{s}}evi{\'c}}},\ }\bibfield  {title} {\bibinfo {title} {Statics and
  dynamics of skyrmions interacting with disorder and nanostructures},\ }\href
  {https://doi.org/10.1103/RevModPhys.94.035005} {\bibfield  {journal}
  {\bibinfo  {journal} {Rev. Mod. Phys.}\ }\textbf {\bibinfo {volume} {94}},\
  \bibinfo {pages} {035005} (\bibinfo {year} {2022})}\BibitemShut {NoStop}%
\bibitem [{\citenamefont {Lauga}\ \emph {et~al.}(2006)\citenamefont {Lauga},
  \citenamefont {DiLuzi}, \citenamefont {Whitesides},\ and\ \citenamefont
  {Stone}}]{lauga2006swimming}%
  \BibitemOpen
  \bibfield  {author} {\bibinfo {author} {\bibfnamefont {E.}~\bibnamefont
  {Lauga}}, \bibinfo {author} {\bibfnamefont {W.~R.}\ \bibnamefont {DiLuzi}},
  \bibinfo {author} {\bibfnamefont {G.~M.}\ \bibnamefont {Whitesides}},\ and\
  \bibinfo {author} {\bibfnamefont {H.~A.}\ \bibnamefont {Stone}},\ }\bibfield
  {title} {\bibinfo {title} {Swimming in circles: {M}otion of bacteria near
  solid boundaries},\ }\href {https://doi.org/10.1529/biophysj.105.069401}
  {\bibfield  {journal} {\bibinfo  {journal} {Biophys. J.}\ }\textbf {\bibinfo
  {volume} {90}},\ \bibinfo {pages} {400} (\bibinfo {year} {2006})}\BibitemShut
  {NoStop}%
\bibitem [{\citenamefont {Lettermann}\ \emph {et~al.}(2026)\citenamefont
  {Lettermann}, \citenamefont {Singer}, \citenamefont {Steinbr{\"u}ck},
  \citenamefont {Ziebert}, \citenamefont {Kanatani}, \citenamefont {Sinnis},
  \citenamefont {Frischknecht},\ and\ \citenamefont
  {Schwarz}}]{lettermann2025chirality}%
  \BibitemOpen
  \bibfield  {author} {\bibinfo {author} {\bibfnamefont {L.}~\bibnamefont
  {Lettermann}}, \bibinfo {author} {\bibfnamefont {M.}~\bibnamefont {Singer}},
  \bibinfo {author} {\bibfnamefont {S.}~\bibnamefont {Steinbr{\"u}ck}},
  \bibinfo {author} {\bibfnamefont {F.}~\bibnamefont {Ziebert}}, \bibinfo
  {author} {\bibfnamefont {S.}~\bibnamefont {Kanatani}}, \bibinfo {author}
  {\bibfnamefont {P.}~\bibnamefont {Sinnis}}, \bibinfo {author} {\bibfnamefont
  {F.}~\bibnamefont {Frischknecht}},\ and\ \bibinfo {author} {\bibfnamefont
  {U.~S.}\ \bibnamefont {Schwarz}},\ }\bibfield  {title} {\bibinfo {title}
  {Chirality of malaria parasites determines their motion patterns},\ }\href
  {https://doi.org/10.1038/s41567-025-03096-0} {\bibfield  {journal} {\bibinfo
  {journal} {Nat. Phys.}\ }\textbf {\bibinfo {volume} {22}},\ \bibinfo {pages}
  {112} (\bibinfo {year} {2026})}\BibitemShut {NoStop}%
\bibitem [{\citenamefont {Hargus}\ \emph {et~al.}(2025)\citenamefont {Hargus},
  \citenamefont {Ghimenti}, \citenamefont {Tailleur},\ and\ \citenamefont {van
  Wijland}}]{hargus2025odd}%
  \BibitemOpen
  \bibfield  {author} {\bibinfo {author} {\bibfnamefont {C.}~\bibnamefont
  {Hargus}}, \bibinfo {author} {\bibfnamefont {F.}~\bibnamefont {Ghimenti}},
  \bibinfo {author} {\bibfnamefont {J.}~\bibnamefont {Tailleur}},\ and\
  \bibinfo {author} {\bibfnamefont {F.}~\bibnamefont {van Wijland}},\
  }\bibfield  {title} {\bibinfo {title} {Odd dynamics of passive objects in a
  chiral active bath},\ }\href {https://doi.org/10.1103/pdpf-sd9q} {\bibfield
  {journal} {\bibinfo  {journal} {Phys. Rev. Lett.}\ }\textbf {\bibinfo
  {volume} {135}},\ \bibinfo {pages} {167102} (\bibinfo {year}
  {2025})}\BibitemShut {NoStop}%
\bibitem [{\citenamefont {W{\'o}jcik}\ and\ \citenamefont
  {Kalz}(2026)}]{wojcik2026chiral}%
  \BibitemOpen
  \bibfield  {author} {\bibinfo {author} {\bibfnamefont {J.}~\bibnamefont
  {W{\'o}jcik}}\ and\ \bibinfo {author} {\bibfnamefont {E.}~\bibnamefont
  {Kalz}},\ }\bibfield  {title} {\bibinfo {title} {The chiral random walk: A
  quantum-inspired framework for odd diffusion},\ }\href
  {https://doi.org/10.1088/1367-2630/aeaf77} {\bibfield  {journal} {\bibinfo
  {journal} {New J. Phys.}\ }\textbf {\bibinfo {volume} {28}},\ \bibinfo
  {pages} {114603} (\bibinfo {year} {2026})}\BibitemShut {NoStop}%
\bibitem [{\citenamefont {Rayleigh}(1892)}]{rayleigh1892obstacles}%
  \BibitemOpen
  \bibfield  {author} {\bibinfo {author} {\bibfnamefont {L.}~\bibnamefont
  {Rayleigh}},\ }\bibfield  {title} {\bibinfo {title} {On the influence of
  obstacles arranged in rectangular order upon the properties of a medium},\
  }\href {https://doi.org/10.1080/14786449208620364} {\bibfield  {journal}
  {\bibinfo  {journal} {Lond. Edinb. Dubl. Phil. Mag.}\ }\textbf {\bibinfo
  {volume} {34}},\ \bibinfo {pages} {481} (\bibinfo {year} {1892})}\BibitemShut
  {NoStop}%
\bibitem [{\citenamefont {Schirmacher}\ \emph {et~al.}(2015)\citenamefont
  {Schirmacher}, \citenamefont {Fuchs}, \citenamefont {H{\"o}fling},\ and\
  \citenamefont {Franosch}}]{schirmacher2015anomalous}%
  \BibitemOpen
  \bibfield  {author} {\bibinfo {author} {\bibfnamefont {W.}~\bibnamefont
  {Schirmacher}}, \bibinfo {author} {\bibfnamefont {B.}~\bibnamefont {Fuchs}},
  \bibinfo {author} {\bibfnamefont {F.}~\bibnamefont {H{\"o}fling}},\ and\
  \bibinfo {author} {\bibfnamefont {T.}~\bibnamefont {Franosch}},\ }\bibfield
  {title} {\bibinfo {title} {Anomalous magnetotransport in disordered
  structures: Classical edge-state percolation},\ }\href
  {https://doi.org/10.1103/PhysRevLett.115.240602} {\bibfield  {journal}
  {\bibinfo  {journal} {Phys. Rev. Lett.}\ }\textbf {\bibinfo {volume} {115}},\
  \bibinfo {pages} {240602} (\bibinfo {year} {2015})}\BibitemShut {NoStop}%
\bibitem [{\citenamefont {Siboni}\ \emph {et~al.}(2018)\citenamefont {Siboni},
  \citenamefont {Schluck}, \citenamefont {Pierz}, \citenamefont {Schumacher},
  \citenamefont {Kazazis}, \citenamefont {Horbach},\ and\ \citenamefont
  {Heinzel}}]{siboni2018nonmonotonic}%
  \BibitemOpen
  \bibfield  {author} {\bibinfo {author} {\bibfnamefont {N.~H.}\ \bibnamefont
  {Siboni}}, \bibinfo {author} {\bibfnamefont {J.}~\bibnamefont {Schluck}},
  \bibinfo {author} {\bibfnamefont {K.}~\bibnamefont {Pierz}}, \bibinfo
  {author} {\bibfnamefont {H.~W.}\ \bibnamefont {Schumacher}}, \bibinfo
  {author} {\bibfnamefont {D.}~\bibnamefont {Kazazis}}, \bibinfo {author}
  {\bibfnamefont {J.}~\bibnamefont {Horbach}},\ and\ \bibinfo {author}
  {\bibfnamefont {T.}~\bibnamefont {Heinzel}},\ }\bibfield  {title} {\bibinfo
  {title} {Nonmonotonic classical magnetoconductivity of a two-dimensional
  electron gas in a disordered array of obstacles},\ }\href
  {https://doi.org/10.1103/PhysRevLett.120.056601} {\bibfield  {journal}
  {\bibinfo  {journal} {Phys. Rev. Lett.}\ }\textbf {\bibinfo {volume} {120}},\
  \bibinfo {pages} {056601} (\bibinfo {year} {2018})}\BibitemShut {NoStop}%
\bibitem [{\citenamefont {Chepizhko}\ and\ \citenamefont
  {Franosch}(2019)}]{chepizhko2019ideal}%
  \BibitemOpen
  \bibfield  {author} {\bibinfo {author} {\bibfnamefont {O.}~\bibnamefont
  {Chepizhko}}\ and\ \bibinfo {author} {\bibfnamefont {T.}~\bibnamefont
  {Franosch}},\ }\bibfield  {title} {\bibinfo {title} {Ideal circle
  microswimmers in crowded media},\ }\href {https://doi.org/10.1039/c8sm02030b}
  {\bibfield  {journal} {\bibinfo  {journal} {Soft Matter}\ }\textbf {\bibinfo
  {volume} {15}},\ \bibinfo {pages} {452} (\bibinfo {year} {2019})}\BibitemShut
  {NoStop}%
\bibitem [{\citenamefont {Chepizhko}\ and\ \citenamefont
  {Franosch}(2020)}]{chepizhko2020random}%
  \BibitemOpen
  \bibfield  {author} {\bibinfo {author} {\bibfnamefont {O.}~\bibnamefont
  {Chepizhko}}\ and\ \bibinfo {author} {\bibfnamefont {T.}~\bibnamefont
  {Franosch}},\ }\bibfield  {title} {\bibinfo {title} {Random motion of a
  circle microswimmer in a random environment},\ }\href
  {https://doi.org/10.1088/1367-2630/ab9708} {\bibfield  {journal} {\bibinfo
  {journal} {New J. Phys.}\ }\textbf {\bibinfo {volume} {22}},\ \bibinfo
  {pages} {073022} (\bibinfo {year} {2020})}\BibitemShut {NoStop}%
\bibitem [{\citenamefont {Uhlig}\ \emph
  {et~al.}(2026{\natexlab{a}})\citenamefont {Uhlig}, \citenamefont
  {W{\'o}jcik}, \citenamefont {Metzler},\ and\ \citenamefont
  {Kalz}}]{uhlig2026chirality}%
  \BibitemOpen
  \bibfield  {author} {\bibinfo {author} {\bibfnamefont {J.}~\bibnamefont
  {Uhlig}}, \bibinfo {author} {\bibfnamefont {J.}~\bibnamefont {W{\'o}jcik}},
  \bibinfo {author} {\bibfnamefont {R.}~\bibnamefont {Metzler}},\ and\ \bibinfo
  {author} {\bibfnamefont {E.}~\bibnamefont {Kalz}},\ }\bibfield  {title}
  {\bibinfo {title} {Chirality enhances diffusion in disordered environments},\
  }\href {https://doi.org/10.1063/5.0352851} {\bibfield  {journal} {\bibinfo
  {journal} {J. Chem. Phys.}\ }\textbf {\bibinfo {volume} {165}},\ \bibinfo
  {pages} {124118} (\bibinfo {year} {2026}{\natexlab{a}})}\BibitemShut
  {NoStop}%
\bibitem [{\citenamefont {Sykes}\ and\ \citenamefont
  {Essam}(1964)}]{sykes1964exact}%
  \BibitemOpen
  \bibfield  {author} {\bibinfo {author} {\bibfnamefont {M.~F.}\ \bibnamefont
  {Sykes}}\ and\ \bibinfo {author} {\bibfnamefont {J.~W.}\ \bibnamefont
  {Essam}},\ }\bibfield  {title} {\bibinfo {title} {Exact critical percolation
  probabilities for site and bond problems in two dimensions},\ }\href
  {https://doi.org/10.1063/1.1704215} {\bibfield  {journal} {\bibinfo
  {journal} {J. Math. Phys.}\ }\textbf {\bibinfo {volume} {5}},\ \bibinfo
  {pages} {1117} (\bibinfo {year} {1964})}\BibitemShut {NoStop}%
\bibitem [{\citenamefont {Saleur}\ and\ \citenamefont
  {Duplantier}(1987)}]{saleur1987exact}%
  \BibitemOpen
  \bibfield  {author} {\bibinfo {author} {\bibfnamefont {H.}~\bibnamefont
  {Saleur}}\ and\ \bibinfo {author} {\bibfnamefont {B.}~\bibnamefont
  {Duplantier}},\ }\bibfield  {title} {\bibinfo {title} {Exact determination of
  the percolation hull exponent in two dimensions},\ }\href
  {https://doi.org/10.1103/PhysRevLett.58.2325} {\bibfield  {journal} {\bibinfo
   {journal} {Phys. Rev. Lett.}\ }\textbf {\bibinfo {volume} {58}},\ \bibinfo
  {pages} {2325} (\bibinfo {year} {1987})}\BibitemShut {NoStop}%
\bibitem [{\citenamefont {Ziff}\ \emph {et~al.}(1984)\citenamefont {Ziff},
  \citenamefont {Cummings},\ and\ \citenamefont {Stells}}]{ziff1984generation}%
  \BibitemOpen
  \bibfield  {author} {\bibinfo {author} {\bibfnamefont {R.~M.}\ \bibnamefont
  {Ziff}}, \bibinfo {author} {\bibfnamefont {P.~T.}\ \bibnamefont {Cummings}},\
  and\ \bibinfo {author} {\bibfnamefont {G.}~\bibnamefont {Stells}},\
  }\bibfield  {title} {\bibinfo {title} {Generation of percolation cluster
  perimeters by a random walk},\ }\href
  {https://doi.org/10.1088/0305-4470/17/15/018} {\bibfield  {journal} {\bibinfo
   {journal} {J. Phys. A}\ }\textbf {\bibinfo {volume} {17}},\ \bibinfo {pages}
  {3009} (\bibinfo {year} {1984})}\BibitemShut {NoStop}%
\bibitem [{\citenamefont {Majid}\ \emph {et~al.}(1984)\citenamefont {Majid},
  \citenamefont {Jan}, \citenamefont {Coniglio},\ and\ \citenamefont
  {Stanley}}]{majid1984kinetic}%
  \BibitemOpen
  \bibfield  {author} {\bibinfo {author} {\bibfnamefont {I.}~\bibnamefont
  {Majid}}, \bibinfo {author} {\bibfnamefont {N.}~\bibnamefont {Jan}}, \bibinfo
  {author} {\bibfnamefont {A.}~\bibnamefont {Coniglio}},\ and\ \bibinfo
  {author} {\bibfnamefont {H.~E.}\ \bibnamefont {Stanley}},\ }\bibfield
  {title} {\bibinfo {title} {Kinetic growth walk: A new model for linear
  polymers},\ }\href {https://doi.org/10.1103/PhysRevLett.52.1257} {\bibfield
  {journal} {\bibinfo  {journal} {Phys. Rev. Lett.}\ }\textbf {\bibinfo
  {volume} {52}},\ \bibinfo {pages} {1257} (\bibinfo {year}
  {1984})}\BibitemShut {NoStop}%
\bibitem [{\citenamefont {Weinrib}\ and\ \citenamefont
  {Trugman}(1985)}]{weinrib1985new}%
  \BibitemOpen
  \bibfield  {author} {\bibinfo {author} {\bibfnamefont {A.}~\bibnamefont
  {Weinrib}}\ and\ \bibinfo {author} {\bibfnamefont {S.~A.}\ \bibnamefont
  {Trugman}},\ }\bibfield  {title} {\bibinfo {title} {A new kinetic walk and
  percolation perimeters},\ }\href {https://doi.org/10.1103/PhysRevB.31.2993}
  {\bibfield  {journal} {\bibinfo  {journal} {Phys. Rev. B}\ }\textbf {\bibinfo
  {volume} {31}},\ \bibinfo {pages} {2993} (\bibinfo {year}
  {1985})}\BibitemShut {NoStop}%
\bibitem [{\citenamefont {Ziff}(1989)}]{ziff1989hull}%
  \BibitemOpen
  \bibfield  {author} {\bibinfo {author} {\bibfnamefont {R.~M.}\ \bibnamefont
  {Ziff}},\ }\bibfield  {title} {\bibinfo {title} {Hull-generating walks},\
  }\href {https://doi.org/https://doi.org/10.1016/0167-2789(89)90222-4}
  {\bibfield  {journal} {\bibinfo  {journal} {Physica D}\ }\textbf {\bibinfo
  {volume} {38}},\ \bibinfo {pages} {377} (\bibinfo {year} {1989})}\BibitemShut
  {NoStop}%
\bibitem [{\citenamefont {Ziff}\ \emph {et~al.}(1991)\citenamefont {Ziff},
  \citenamefont {Kong},\ and\ \citenamefont {Cohen}}]{ziff1991lorentz}%
  \BibitemOpen
  \bibfield  {author} {\bibinfo {author} {\bibfnamefont {R.~M.}\ \bibnamefont
  {Ziff}}, \bibinfo {author} {\bibfnamefont {X.~P.}\ \bibnamefont {Kong}},\
  and\ \bibinfo {author} {\bibfnamefont {E.~G.~D.}\ \bibnamefont {Cohen}},\
  }\bibfield  {title} {\bibinfo {title} {Lorentz lattice-gas and kinetic-walk
  model},\ }\href {https://doi.org/10.1103/PhysRevA.44.2410} {\bibfield
  {journal} {\bibinfo  {journal} {Phys. Rev. A}\ }\textbf {\bibinfo {volume}
  {44}},\ \bibinfo {pages} {2410} (\bibinfo {year} {1991})}\BibitemShut
  {NoStop}%
\bibitem [{\citenamefont {Newman}\ and\ \citenamefont
  {Ziff}(2000)}]{newman2000efficient}%
  \BibitemOpen
  \bibfield  {author} {\bibinfo {author} {\bibfnamefont {M.~E.~J.}\
  \bibnamefont {Newman}}\ and\ \bibinfo {author} {\bibfnamefont {R.~M.}\
  \bibnamefont {Ziff}},\ }\bibfield  {title} {\bibinfo {title} {Efficient
  {M}onte {C}arlo algorithm and high-precision results for percolation},\
  }\href {https://doi.org/10.1103/PhysRevLett.85.4104} {\bibfield  {journal}
  {\bibinfo  {journal} {Phys. Rev. Lett.}\ }\textbf {\bibinfo {volume} {85}},\
  \bibinfo {pages} {4104} (\bibinfo {year} {2000})}\BibitemShut {NoStop}%
\bibitem [{\citenamefont {Ziff}(2011)}]{ziff2011results}%
  \BibitemOpen
  \bibfield  {author} {\bibinfo {author} {\bibfnamefont {R.~M.}\ \bibnamefont
  {Ziff}},\ }\bibfield  {title} {\bibinfo {title} {Results for a critical
  threshold, the correction-to-scaling exponent and susceptibility amplitude
  ratio for 2d percolation},\ }\href
  {https://doi.org/10.1016/j.phpro.2011.06.009} {\bibfield  {journal} {\bibinfo
   {journal} {Phys. Procedia}\ }\textbf {\bibinfo {volume} {15}},\ \bibinfo
  {pages} {12} (\bibinfo {year} {2011})}\BibitemShut {NoStop}%
\bibitem [{\citenamefont {Jacobsen}(2015)}]{jacobsen2015critical}%
  \BibitemOpen
  \bibfield  {author} {\bibinfo {author} {\bibfnamefont {J.~L.}\ \bibnamefont
  {Jacobsen}},\ }\bibfield  {title} {\bibinfo {title} {Critical points of
  {P}otts and {O}({N}) models from eigenvalue identities in periodic
  {T}emperley--{L}ieb algebras},\ }\href
  {https://doi.org/10.1088/1751-8113/48/45/454003} {\bibfield  {journal}
  {\bibinfo  {journal} {J. Phys. A}\ }\textbf {\bibinfo {volume} {48}},\
  \bibinfo {pages} {454003} (\bibinfo {year} {2015})}\BibitemShut {NoStop}%
\bibitem [{\citenamefont {Mertens}(2022)}]{mertens2022exact}%
  \BibitemOpen
  \bibfield  {author} {\bibinfo {author} {\bibfnamefont {S.}~\bibnamefont
  {Mertens}},\ }\bibfield  {title} {\bibinfo {title} {Exact site-percolation
  probability on the square lattice},\ }\href
  {https://doi.org/10.1088/1751-8121/ac4195} {\bibfield  {journal} {\bibinfo
  {journal} {J. Phys. A}\ }\textbf {\bibinfo {volume} {55}},\ \bibinfo {pages}
  {334002} (\bibinfo {year} {2022})}\BibitemShut {NoStop}%
\bibitem [{\citenamefont {Hoshen}\ and\ \citenamefont
  {Kopelman}(1976)}]{hoshen1976percolation}%
  \BibitemOpen
  \bibfield  {author} {\bibinfo {author} {\bibfnamefont {J.}~\bibnamefont
  {Hoshen}}\ and\ \bibinfo {author} {\bibfnamefont {R.}~\bibnamefont
  {Kopelman}},\ }\bibfield  {title} {\bibinfo {title} {Percolation and cluster
  distribution. {I}. {C}luster multiple labeling technique and critical
  concentration algorithm},\ }\href {https://doi.org/10.1103/PhysRevB.14.3438}
  {\bibfield  {journal} {\bibinfo  {journal} {Phys. Rev. B}\ }\textbf {\bibinfo
  {volume} {14}},\ \bibinfo {pages} {3438} (\bibinfo {year}
  {1976})}\BibitemShut {NoStop}%
\bibitem [{\citenamefont {Condamin}\ \emph {et~al.}(2007)\citenamefont
  {Condamin}, \citenamefont {B\'enichou},\ and\ \citenamefont
  {Klafter}}]{condamin2007firstprl}%
  \BibitemOpen
  \bibfield  {author} {\bibinfo {author} {\bibfnamefont {S.}~\bibnamefont
  {Condamin}}, \bibinfo {author} {\bibfnamefont {O.}~\bibnamefont
  {B\'enichou}},\ and\ \bibinfo {author} {\bibfnamefont {J.}~\bibnamefont
  {Klafter}},\ }\bibfield  {title} {\bibinfo {title} {First-passage time
  distributions for subdiffusion in confined geometry},\ }\href
  {https://doi.org/10.1103/PhysRevLett.98.250602} {\bibfield  {journal}
  {\bibinfo  {journal} {Phys. Rev. Lett.}\ }\textbf {\bibinfo {volume} {98}},\
  \bibinfo {pages} {250602} (\bibinfo {year} {2007})}\BibitemShut {NoStop}%
\bibitem [{\citenamefont {Feder}(1988)}]{feder1988fractals}%
  \BibitemOpen
  \bibfield  {author} {\bibinfo {author} {\bibfnamefont {J.}~\bibnamefont
  {Feder}},\ }\href {https://doi.org/10.1007/978-1-4899-2124-6} {\emph
  {\bibinfo {title} {Fractals}}}\ (\bibinfo  {publisher} {Plenum Press},\
  \bibinfo {address} {New York, NY},\ \bibinfo {year} {1988})\BibitemShut
  {NoStop}%
\bibitem [{\citenamefont {Falconer}(2003)}]{falconer2003fractal}%
  \BibitemOpen
  \bibfield  {author} {\bibinfo {author} {\bibfnamefont {K.}~\bibnamefont
  {Falconer}},\ }\href {https://doi.org/10.1002/0470013850} {\emph {\bibinfo
  {title} {Fractal Geometry: Mathematical Foundations and Applications}}},\
  \bibinfo {edition} {2nd}\ ed.\ (\bibinfo  {publisher} {Wiley},\ \bibinfo
  {address} {Chichester},\ \bibinfo {year} {2003})\BibitemShut {NoStop}%
\bibitem [{\citenamefont {Ziff}(1986)}]{ziff1986test}%
  \BibitemOpen
  \bibfield  {author} {\bibinfo {author} {\bibfnamefont {R.~M.}\ \bibnamefont
  {Ziff}},\ }\bibfield  {title} {\bibinfo {title} {Test of scaling exponents
  for percolation-cluster perimeters},\ }\href
  {https://doi.org/10.1103/PhysRevLett.56.545} {\bibfield  {journal} {\bibinfo
  {journal} {Phys. Rev. Lett.}\ }\textbf {\bibinfo {volume} {56}},\ \bibinfo
  {pages} {545} (\bibinfo {year} {1986})}\BibitemShut {NoStop}%
\bibitem [{\citenamefont {Kac}(1966)}]{kac1966can}%
  \BibitemOpen
  \bibfield  {author} {\bibinfo {author} {\bibfnamefont {M.}~\bibnamefont
  {Kac}},\ }\bibfield  {title} {\bibinfo {title} {Can one hear the shape of a
  drum?},\ }\href {https://doi.org/10.1080/00029890.1966.11970915} {\bibfield
  {journal} {\bibinfo  {journal} {Am. Math. Mon.}\ }\textbf {\bibinfo {volume}
  {73}},\ \bibinfo {pages} {1} (\bibinfo {year} {1966})}\BibitemShut {NoStop}%
\bibitem [{\citenamefont {Lapidus}(1991{\natexlab{a}})}]{lapidus1991can}%
  \BibitemOpen
  \bibfield  {author} {\bibinfo {author} {\bibfnamefont {M.~L.}\ \bibnamefont
  {Lapidus}},\ }\bibfield  {title} {\bibinfo {title} {Can one hear the shape of
  a fractal drum? {P}artial resolution of the {W}eyl--{B}erry conjecture},\
  }in\ \href {https://doi.org/doi.org/10.1007/978-1-4613-9711-3_13} {\emph
  {\bibinfo {booktitle} {Geometric Analysis and Computer Graphics}}},\ \bibinfo
  {series} {Math. Sci. Res. Inst. Publ.}, Vol.~\bibinfo {volume} {17},\
  \bibinfo {editor} {edited by\ \bibinfo {editor} {\bibfnamefont
  {P.}~\bibnamefont {Concus}}, \bibinfo {editor} {\bibfnamefont
  {R.}~\bibnamefont {Finn}},\ and\ \bibinfo {editor} {\bibfnamefont {D.~A.}\
  \bibnamefont {Hoffman}}}\ (\bibinfo  {publisher} {Springer},\ \bibinfo
  {address} {New York},\ \bibinfo {year} {1991})\ pp.\ \bibinfo {pages}
  {119--126}\BibitemShut {NoStop}%
\bibitem [{\citenamefont {Lapidus}(1991{\natexlab{b}})}]{lapidus1991fractal}%
  \BibitemOpen
  \bibfield  {author} {\bibinfo {author} {\bibfnamefont {M.~L.}\ \bibnamefont
  {Lapidus}},\ }\bibfield  {title} {\bibinfo {title} {Fractal drum, inverse
  spectral problems for elliptic operators and a partial resolution of the
  {W}eyl--{B}erry conjecture},\ }\href
  {https://doi.org/doi.org/10.1090/S0002-9947-1991-0994168-5} {\bibfield
  {journal} {\bibinfo  {journal} {Trans. Amer. Math. Soc.}\ }\textbf {\bibinfo
  {volume} {325}},\ \bibinfo {pages} {465} (\bibinfo {year}
  {1991}{\natexlab{b}})}\BibitemShut {NoStop}%
\bibitem [{\citenamefont {Eisenmann}\ \emph {et~al.}(2026)\citenamefont
  {Eisenmann}, \citenamefont {Cosme}, \citenamefont {van Sterkenburg},
  \citenamefont {Franca}, \citenamefont {Seguy}, \citenamefont
  {Tatulea-Codrean},\ and\ \citenamefont {Jalaal}}]{eisenmann2026chirality}%
  \BibitemOpen
  \bibfield  {author} {\bibinfo {author} {\bibfnamefont {I.}~\bibnamefont
  {Eisenmann}}, \bibinfo {author} {\bibfnamefont {P.}~\bibnamefont {Cosme}},
  \bibinfo {author} {\bibfnamefont {H.}~\bibnamefont {van Sterkenburg}},
  \bibinfo {author} {\bibfnamefont {H.}~\bibnamefont {Franca}}, \bibinfo
  {author} {\bibfnamefont {T.}~\bibnamefont {Seguy}}, \bibinfo {author}
  {\bibfnamefont {M.}~\bibnamefont {Tatulea-Codrean}},\ and\ \bibinfo {author}
  {\bibfnamefont {M.}~\bibnamefont {Jalaal}},\ }\bibfield  {title} {\bibinfo
  {title} {Chirality inversion and odd transport in a suspension of living
  rotors},\ }\bibfield  {journal} {\bibinfo  {journal} {E-print}\ }\href
  {https://doi.org/10.48550/arXiv.2610.41016} {10.48550/arXiv.2610.41016}
  (\bibinfo {year} {2026})\BibitemShut {NoStop}%
\bibitem [{\citenamefont {Keitt}\ \emph {et~al.}(1997)\citenamefont {Keitt},
  \citenamefont {Urban},\ and\ \citenamefont {Milne}}]{keitt1997detecting}%
  \BibitemOpen
  \bibfield  {author} {\bibinfo {author} {\bibfnamefont {T.}~\bibnamefont
  {Keitt}}, \bibinfo {author} {\bibfnamefont {D.}~\bibnamefont {Urban}},\ and\
  \bibinfo {author} {\bibfnamefont {B.}~\bibnamefont {Milne}},\ }\bibfield
  {title} {\bibinfo {title} {Detecting critical scales in fragmented
  landscapes},\ }\href {https://doi.org/10.5751/ES-00015-010104} {\bibfield
  {journal} {\bibinfo  {journal} {Conserv. Ecol.}\ }\textbf {\bibinfo {volume}
  {1}},\ \bibinfo {pages} {4} (\bibinfo {year} {1997})}\BibitemShut {NoStop}%
\bibitem [{\citenamefont {Malanson}\ and\ \citenamefont
  {Cramer}(1999)}]{malanson1999ants}%
  \BibitemOpen
  \bibfield  {author} {\bibinfo {author} {\bibfnamefont {G.~P.}\ \bibnamefont
  {Malanson}}\ and\ \bibinfo {author} {\bibfnamefont {B.~E.}\ \bibnamefont
  {Cramer}},\ }\bibfield  {title} {\bibinfo {title} {Ants in labyrinths:
  Lessons for critical landscapes},\ }\href
  {https://doi.org/https://doi.org/10.1111/0033-0124.00154} {\bibfield
  {journal} {\bibinfo  {journal} {Prof. Geogr.}\ }\textbf {\bibinfo {volume}
  {51}},\ \bibinfo {pages} {155} (\bibinfo {year} {1999})}\BibitemShut
  {NoStop}%
\bibitem [{\citenamefont {Priezzhev}\ \emph {et~al.}(1996)\citenamefont
  {Priezzhev}, \citenamefont {Dhar}, \citenamefont {Dhar},\ and\ \citenamefont
  {Krishnamurthy}}]{priezzhev1996eulerian}%
  \BibitemOpen
  \bibfield  {author} {\bibinfo {author} {\bibfnamefont {V.~B.}\ \bibnamefont
  {Priezzhev}}, \bibinfo {author} {\bibfnamefont {D.}~\bibnamefont {Dhar}},
  \bibinfo {author} {\bibfnamefont {A.}~\bibnamefont {Dhar}},\ and\ \bibinfo
  {author} {\bibfnamefont {S.}~\bibnamefont {Krishnamurthy}},\ }\bibfield
  {title} {\bibinfo {title} {Eulerian walkers as a model of self-organized
  criticality},\ }\href {https://doi.org/10.1103/PhysRevLett.77.5079}
  {\bibfield  {journal} {\bibinfo  {journal} {Phys. Rev. Lett.}\ }\textbf
  {\bibinfo {volume} {77}},\ \bibinfo {pages} {5079} (\bibinfo {year}
  {1996})}\BibitemShut {NoStop}%
\bibitem [{\citenamefont {Yanovski}\ \emph {et~al.}(2003)\citenamefont
  {Yanovski}, \citenamefont {Wagner},\ and\ \citenamefont
  {Bruckstein}}]{yanovski2003distributed}%
  \BibitemOpen
  \bibfield  {author} {\bibinfo {author} {\bibfnamefont {V.}~\bibnamefont
  {Yanovski}}, \bibinfo {author} {\bibfnamefont {I.}~\bibnamefont {Wagner}},\
  and\ \bibinfo {author} {\bibfnamefont {A.}~\bibnamefont {Bruckstein}},\
  }\bibfield  {title} {\bibinfo {title} {A distributed ant algorithm for
  efficiently patrolling a network},\ }\href
  {https://doi.org/10.1007/s00453-003-1030-9} {\bibfield  {journal} {\bibinfo
  {journal} {Algorithmica}\ }\textbf {\bibinfo {volume} {37}},\ \bibinfo
  {pages} {165} (\bibinfo {year} {2003})}\BibitemShut {NoStop}%
\bibitem [{\citenamefont {Skinner}\ \emph {et~al.}(2013)\citenamefont
  {Skinner}, \citenamefont {Schnyder}, \citenamefont {Aarts}, \citenamefont
  {Horbach},\ and\ \citenamefont {Dullens}}]{skinner2013localization}%
  \BibitemOpen
  \bibfield  {author} {\bibinfo {author} {\bibfnamefont {T.~O.~E.}\
  \bibnamefont {Skinner}}, \bibinfo {author} {\bibfnamefont {S.~K.}\
  \bibnamefont {Schnyder}}, \bibinfo {author} {\bibfnamefont {D.~G. A.~L.}\
  \bibnamefont {Aarts}}, \bibinfo {author} {\bibfnamefont {J.}~\bibnamefont
  {Horbach}},\ and\ \bibinfo {author} {\bibfnamefont {R.~P.~A.}\ \bibnamefont
  {Dullens}},\ }\bibfield  {title} {\bibinfo {title} {Localization dynamics of
  fluids in random confinement},\ }\href
  {https://doi.org/10.1103/PhysRevLett.111.128301} {\bibfield  {journal}
  {\bibinfo  {journal} {Phys. Rev. Lett.}\ }\textbf {\bibinfo {volume} {111}},\
  \bibinfo {pages} {128301} (\bibinfo {year} {2013})}\BibitemShut {NoStop}%
\bibitem [{\citenamefont {Chan}\ \emph {et~al.}(2024)\citenamefont {Chan},
  \citenamefont {Wu}, \citenamefont {Qiao}, \citenamefont {Fong}, \citenamefont
  {Yang}, \citenamefont {Han},\ and\ \citenamefont {Zhang}}]{chan2024chiral}%
  \BibitemOpen
  \bibfield  {author} {\bibinfo {author} {\bibfnamefont {C.~W.}\ \bibnamefont
  {Chan}}, \bibinfo {author} {\bibfnamefont {D.}~\bibnamefont {Wu}}, \bibinfo
  {author} {\bibfnamefont {K.}~\bibnamefont {Qiao}}, \bibinfo {author}
  {\bibfnamefont {K.~L.}\ \bibnamefont {Fong}}, \bibinfo {author}
  {\bibfnamefont {Z.}~\bibnamefont {Yang}}, \bibinfo {author} {\bibfnamefont
  {Y.}~\bibnamefont {Han}},\ and\ \bibinfo {author} {\bibfnamefont
  {R.}~\bibnamefont {Zhang}},\ }\bibfield  {title} {\bibinfo {title} {Chiral
  active particles are sensitive reporters to environmental geometry},\ }\href
  {https://doi.org/10.1038/s41467-024-45531-5} {\bibfield  {journal} {\bibinfo
  {journal} {Nat. Commun.}\ }\textbf {\bibinfo {volume} {15}},\ \bibinfo
  {pages} {1406} (\bibinfo {year} {2024})}\BibitemShut {NoStop}%
\bibitem [{\citenamefont {Bhattacharjee}\ and\ \citenamefont
  {Datta}(2019)}]{bhattacharjee2019bacterial}%
  \BibitemOpen
  \bibfield  {author} {\bibinfo {author} {\bibfnamefont {T.}~\bibnamefont
  {Bhattacharjee}}\ and\ \bibinfo {author} {\bibfnamefont {S.~S.}\ \bibnamefont
  {Datta}},\ }\bibfield  {title} {\bibinfo {title} {Bacterial hopping and
  trapping in porous media},\ }\href {https://doi.org/10.1038/s41467-019-10115-1}
  {\bibfield  {journal} {\bibinfo  {journal} {Nat. Commun.}\ }\textbf {\bibinfo
  {volume} {10}},\ \bibinfo {pages} {2075} (\bibinfo {year}
  {2019})}\BibitemShut {NoStop}%
\bibitem [{\citenamefont {Sch{\"u}tte}\ \emph {et~al.}(2014)\citenamefont
  {Sch{\"u}tte}, \citenamefont {Iwasaki}, \citenamefont {Rosch},\ and\
  \citenamefont {Nagaosa}}]{schutte2014inertia}%
  \BibitemOpen
  \bibfield  {author} {\bibinfo {author} {\bibfnamefont {C.}~\bibnamefont
  {Sch{\"u}tte}}, \bibinfo {author} {\bibfnamefont {J.}~\bibnamefont
  {Iwasaki}}, \bibinfo {author} {\bibfnamefont {A.}~\bibnamefont {Rosch}},\
  and\ \bibinfo {author} {\bibfnamefont {N.}~\bibnamefont {Nagaosa}},\
  }\bibfield  {title} {\bibinfo {title} {Inertia, diffusion, and dynamics of a
  driven skyrmion},\ }\href {https://doi.org/10.1103/PhysRevB.90.174434}
  {\bibfield  {journal} {\bibinfo  {journal} {Phys. Rev. B}\ }\textbf {\bibinfo
  {volume} {90}},\ \bibinfo {pages} {174434} (\bibinfo {year}
  {2014})}\BibitemShut {NoStop}%
\bibitem [{\citenamefont {Rieger}\ \emph {et~al.}(2023)\citenamefont {Rieger},
  \citenamefont {Wei{\ss{}}enhofer},\ and\ \citenamefont
  {Nowak}}]{rieger2023defect}%
  \BibitemOpen
  \bibfield  {author} {\bibinfo {author} {\bibfnamefont {P.}~\bibnamefont
  {Rieger}}, \bibinfo {author} {\bibfnamefont {M.}~\bibnamefont
  {Wei{\ss{}}enhofer}},\ and\ \bibinfo {author} {\bibfnamefont
  {U.}~\bibnamefont {Nowak}},\ }\bibfield  {title} {\bibinfo {title}
  {Defect-enhanced diffusion of magnetic skyrmions},\ }\href
  {https://doi.org/10.1103/PhysRevB.108.144417} {\bibfield  {journal} {\bibinfo
   {journal} {Phys. Rev. B}\ }\textbf {\bibinfo {volume} {108}},\ \bibinfo
  {pages} {144417} (\bibinfo {year} {2023})}\BibitemShut {NoStop}%
\bibitem [{\citenamefont {Brems}\ \emph {et~al.}(2025)\citenamefont {Brems},
  \citenamefont {Sparmann}, \citenamefont {Fr{\"o}hlich}, \citenamefont {Dany},
  \citenamefont {Roth{\"o}rl}, \citenamefont {Kammerbauer}, \citenamefont
  {Jefremovas}, \citenamefont {Farago}, \citenamefont {Kl{\"a}ui},\ and\
  \citenamefont {Virnau}}]{brems2025realizing}%
  \BibitemOpen
  \bibfield  {author} {\bibinfo {author} {\bibfnamefont {M.~A.}\ \bibnamefont
  {Brems}}, \bibinfo {author} {\bibfnamefont {T.}~\bibnamefont {Sparmann}},
  \bibinfo {author} {\bibfnamefont {S.~M.}\ \bibnamefont {Fr{\"o}hlich}},
  \bibinfo {author} {\bibfnamefont {L.-C.}\ \bibnamefont {Dany}}, \bibinfo
  {author} {\bibfnamefont {J.}~\bibnamefont {Roth{\"o}rl}}, \bibinfo {author}
  {\bibfnamefont {F.}~\bibnamefont {Kammerbauer}}, \bibinfo {author}
  {\bibfnamefont {E.~M.}\ \bibnamefont {Jefremovas}}, \bibinfo {author}
  {\bibfnamefont {O.}~\bibnamefont {Farago}}, \bibinfo {author} {\bibfnamefont
  {M.}~\bibnamefont {Kl{\"a}ui}},\ and\ \bibinfo {author} {\bibfnamefont
  {P.}~\bibnamefont {Virnau}},\ }\bibfield  {title} {\bibinfo {title}
  {Realizing quantitative quasiparticle modeling of skyrmion dynamics in
  arbitrary potentials},\ }\href
  {https://doi.org/10.1103/PhysRevLett.134.046701} {\bibfield  {journal}
  {\bibinfo  {journal} {Phys. Rev. Lett.}\ }\textbf {\bibinfo {volume} {134}},\
  \bibinfo {pages} {046701} (\bibinfo {year} {2025})}\BibitemShut {NoStop}%
\bibitem [{\citenamefont {Kern}\ \emph {et~al.}(2022)\citenamefont {Kern},
  \citenamefont {Pfau}, \citenamefont {Deinhart}, \citenamefont {Schneider},
  \citenamefont {Klose}, \citenamefont {Gerlinger}, \citenamefont {Wittrock},
  \citenamefont {Engel}, \citenamefont {Will}, \citenamefont {G{\"u}nther},
  \citenamefont {Liefferink}, \citenamefont {Mentink}, \citenamefont {Wintz},
  \citenamefont {Weigand}, \citenamefont {Huang}, \citenamefont {Battistelli},
  \citenamefont {Metternich}, \citenamefont {B{\"u}ttner}, \citenamefont
  {H{\"o}flich},\ and\ \citenamefont {Eisebitt}}]{kern2022deterministic}%
  \BibitemOpen
  \bibfield  {author} {\bibinfo {author} {\bibfnamefont {L.-M.}\ \bibnamefont
  {Kern}}, \bibinfo {author} {\bibfnamefont {B.}~\bibnamefont {Pfau}}, \bibinfo
  {author} {\bibfnamefont {V.}~\bibnamefont {Deinhart}}, \bibinfo {author}
  {\bibfnamefont {M.}~\bibnamefont {Schneider}}, \bibinfo {author}
  {\bibfnamefont {C.}~\bibnamefont {Klose}}, \bibinfo {author} {\bibfnamefont
  {K.}~\bibnamefont {Gerlinger}}, \bibinfo {author} {\bibfnamefont
  {S.}~\bibnamefont {Wittrock}}, \bibinfo {author} {\bibfnamefont
  {D.}~\bibnamefont {Engel}}, \bibinfo {author} {\bibfnamefont
  {I.}~\bibnamefont {Will}}, \bibinfo {author} {\bibfnamefont {C.~M.}\
  \bibnamefont {G{\"u}nther}}, \bibinfo {author} {\bibfnamefont
  {R.}~\bibnamefont {Liefferink}}, \bibinfo {author} {\bibfnamefont {J.~H.}\
  \bibnamefont {Mentink}}, \bibinfo {author} {\bibfnamefont {S.}~\bibnamefont
  {Wintz}}, \bibinfo {author} {\bibfnamefont {M.}~\bibnamefont {Weigand}},
  \bibinfo {author} {\bibfnamefont {M.-J.}\ \bibnamefont {Huang}}, \bibinfo
  {author} {\bibfnamefont {R.}~\bibnamefont {Battistelli}}, \bibinfo {author}
  {\bibfnamefont {D.}~\bibnamefont {Metternich}}, \bibinfo {author}
  {\bibfnamefont {F.}~\bibnamefont {B{\"u}ttner}}, \bibinfo {author}
  {\bibfnamefont {K.}~\bibnamefont {H{\"o}flich}},\ and\ \bibinfo {author}
  {\bibfnamefont {S.}~\bibnamefont {Eisebitt}},\ }\bibfield  {title} {\bibinfo
  {title} {Deterministic generation and guided motion of magnetic skyrmions by
  focused {He}+-ion irradiation},\ }\href
  {https://doi.org/10.1021/acs.nanolett.2c00670} {\bibfield  {journal}
  {\bibinfo  {journal} {Nano Lett.}\ }\textbf {\bibinfo {volume} {22}},\
  \bibinfo {pages} {4028} (\bibinfo {year} {2022})}\BibitemShut {NoStop}%
\bibitem [{\citenamefont {Kalz}\ \emph {et~al.}(2022)\citenamefont {Kalz},
  \citenamefont {Vuijk}, \citenamefont {Abdoli}, \citenamefont {Sommer},
  \citenamefont {L{\"o}wen},\ and\ \citenamefont
  {Sharma}}]{kalz2022collisions}%
  \BibitemOpen
  \bibfield  {author} {\bibinfo {author} {\bibfnamefont {E.}~\bibnamefont
  {Kalz}}, \bibinfo {author} {\bibfnamefont {H.~D.}\ \bibnamefont {Vuijk}},
  \bibinfo {author} {\bibfnamefont {I.}~\bibnamefont {Abdoli}}, \bibinfo
  {author} {\bibfnamefont {J.-U.}\ \bibnamefont {Sommer}}, \bibinfo {author}
  {\bibfnamefont {H.}~\bibnamefont {L{\"o}wen}},\ and\ \bibinfo {author}
  {\bibfnamefont {A.}~\bibnamefont {Sharma}},\ }\bibfield  {title} {\bibinfo
  {title} {Collisions enhance self-diffusion in odd-diffusive systems},\ }\href
  {https://doi.org/10.1103/PhysRevLett.129.090601} {\bibfield  {journal}
  {\bibinfo  {journal} {Phys. Rev. Lett.}\ }\textbf {\bibinfo {volume} {129}},\
  \bibinfo {pages} {090601} (\bibinfo {year} {2022})}\BibitemShut {NoStop}%
\bibitem [{\citenamefont {He}\ \emph {et~al.}(2008)\citenamefont {He},
  \citenamefont {Burov}, \citenamefont {Metzler},\ and\ \citenamefont
  {Barkai}}]{he2008random}%
  \BibitemOpen
  \bibfield  {author} {\bibinfo {author} {\bibfnamefont {Y.}~\bibnamefont
  {He}}, \bibinfo {author} {\bibfnamefont {S.}~\bibnamefont {Burov}}, \bibinfo
  {author} {\bibfnamefont {R.}~\bibnamefont {Metzler}},\ and\ \bibinfo {author}
  {\bibfnamefont {E.}~\bibnamefont {Barkai}},\ }\bibfield  {title} {\bibinfo
  {title} {Random time-scale invariant diffusion and transport coefficients},\
  }\href {https://doi.org/10.1103/PhysRevLett.101.058101} {\bibfield  {journal}
  {\bibinfo  {journal} {Phys. Rev. Lett.}\ }\textbf {\bibinfo {volume} {101}},\
  \bibinfo {pages} {058101} (\bibinfo {year} {2008})}\BibitemShut {NoStop}%
\bibitem [{\citenamefont {Metzler}\ \emph {et~al.}(2014)\citenamefont
  {Metzler}, \citenamefont {Jeon}, \citenamefont {Cherstvy},\ and\
  \citenamefont {Barkai}}]{metzler2014anomalous}%
  \BibitemOpen
  \bibfield  {author} {\bibinfo {author} {\bibfnamefont {R.}~\bibnamefont
  {Metzler}}, \bibinfo {author} {\bibfnamefont {J.-H.}\ \bibnamefont {Jeon}},
  \bibinfo {author} {\bibfnamefont {A.~G.}\ \bibnamefont {Cherstvy}},\ and\
  \bibinfo {author} {\bibfnamefont {E.}~\bibnamefont {Barkai}},\ }\bibfield
  {title} {\bibinfo {title} {Anomalous diffusion models and their properties:
  {N}on-stationarity, non-ergodicity, and ageing at the centenary of single
  particle tracking},\ }\href {https://doi.org/10.1039/c4cp03465a} {\bibfield
  {journal} {\bibinfo  {journal} {Phys. Chem. Chem. Phys.}\ }\textbf {\bibinfo
  {volume} {16}},\ \bibinfo {pages} {24128} (\bibinfo {year}
  {2014})}\BibitemShut {NoStop}%
\bibitem [{\citenamefont {Meroz}\ \emph {et~al.}(2010)\citenamefont {Meroz},
  \citenamefont {Sokolov},\ and\ \citenamefont
  {Klafter}}]{meroz2010subdiffusion}%
  \BibitemOpen
  \bibfield  {author} {\bibinfo {author} {\bibfnamefont {Y.}~\bibnamefont
  {Meroz}}, \bibinfo {author} {\bibfnamefont {I.~M.}\ \bibnamefont {Sokolov}},\
  and\ \bibinfo {author} {\bibfnamefont {J.}~\bibnamefont {Klafter}},\
  }\bibfield  {title} {\bibinfo {title} {Subdiffusion of mixed origins: When
  ergodicity and nonergodicity coexist},\ }\href
  {https://doi.org/10.1103/PhysRevE.81.010101} {\bibfield  {journal} {\bibinfo
  {journal} {Phys. Rev. E}\ }\textbf {\bibinfo {volume} {81}},\ \bibinfo
  {pages} {010101(R)} (\bibinfo {year} {2010})}\BibitemShut {NoStop}%
\bibitem [{\citenamefont {Meroz}\ \emph {et~al.}(2013)\citenamefont {Meroz},
  \citenamefont {Sokolov},\ and\ \citenamefont {Klafter}}]{meroz2013test}%
  \BibitemOpen
  \bibfield  {author} {\bibinfo {author} {\bibfnamefont {Y.}~\bibnamefont
  {Meroz}}, \bibinfo {author} {\bibfnamefont {I.~M.}\ \bibnamefont {Sokolov}},\
  and\ \bibinfo {author} {\bibfnamefont {J.}~\bibnamefont {Klafter}},\
  }\bibfield  {title} {\bibinfo {title} {Test for determining a subdiffusive
  model in ergodic systems from single trajectories},\ }\href
  {https://doi.org/10.1103/PhysRevLett.110.090601} {\bibfield  {journal}
  {\bibinfo  {journal} {Phys. Rev. Lett.}\ }\textbf {\bibinfo {volume} {110}},\
  \bibinfo {pages} {090601} (\bibinfo {year} {2013})}\BibitemShut {NoStop}%
\bibitem [{\citenamefont {Cherstvy}\ \emph {et~al.}(2018)\citenamefont
  {Cherstvy}, \citenamefont {Thapa}, \citenamefont {Mardoukhi}, \citenamefont
  {Chechkin},\ and\ \citenamefont {Metzler}}]{cherstvy2018time}%
  \BibitemOpen
  \bibfield  {author} {\bibinfo {author} {\bibfnamefont {A.~G.}\ \bibnamefont
  {Cherstvy}}, \bibinfo {author} {\bibfnamefont {S.}~\bibnamefont {Thapa}},
  \bibinfo {author} {\bibfnamefont {Y.}~\bibnamefont {Mardoukhi}}, \bibinfo
  {author} {\bibfnamefont {A.~V.}\ \bibnamefont {Chechkin}},\ and\ \bibinfo
  {author} {\bibfnamefont {R.}~\bibnamefont {Metzler}},\ }\bibfield  {title}
  {\bibinfo {title} {Time averages and their statistical variation for the
  {O}rnstein-{U}hlenbeck process: Role of initial particle distributions and
  relaxation to stationarity},\ }\href
  {https://doi.org/10.1103/PhysRevE.98.022134} {\bibfield  {journal} {\bibinfo
  {journal} {Phys. Rev. E}\ }\textbf {\bibinfo {volume} {98}},\ \bibinfo
  {pages} {022134} (\bibinfo {year} {2018})}\BibitemShut {NoStop}%
\bibitem [{\citenamefont {Mardoukhi}\ \emph {et~al.}(2015)\citenamefont
  {Mardoukhi}, \citenamefont {Jeon},\ and\ \citenamefont
  {Metzler}}]{mardoukhi2015geometry}%
  \BibitemOpen
  \bibfield  {author} {\bibinfo {author} {\bibfnamefont {Y.}~\bibnamefont
  {Mardoukhi}}, \bibinfo {author} {\bibfnamefont {J.-H.}\ \bibnamefont
  {Jeon}},\ and\ \bibinfo {author} {\bibfnamefont {R.}~\bibnamefont
  {Metzler}},\ }\bibfield  {title} {\bibinfo {title} {Geometry controlled
  anomalous diffusion in random fractal geometries: {L}ooking beyond the
  infinite cluster},\ }\href {https://doi.org/10.1039/c5cp03548a} {\bibfield
  {journal} {\bibinfo  {journal} {Phys. Chem. Chem. Phys.}\ }\textbf {\bibinfo
  {volume} {17}},\ \bibinfo {pages} {30134} (\bibinfo {year}
  {2015})}\BibitemShut {NoStop}%
\bibitem [{\citenamefont {Rammal}\ and\ \citenamefont
  {Toulouse}(1983)}]{rammal1983random}%
  \BibitemOpen
  \bibfield  {author} {\bibinfo {author} {\bibfnamefont {R.}~\bibnamefont
  {Rammal}}\ and\ \bibinfo {author} {\bibfnamefont {G.}~\bibnamefont
  {Toulouse}},\ }\bibfield  {title} {\bibinfo {title} {Random walks on fractal
  structures and percolation clusters},\ }\href
  {https://doi.org/10.1051/jphyslet:0198300440101300} {\bibfield  {journal}
  {\bibinfo  {journal} {J. Physique Lett.}\ }\textbf {\bibinfo {volume} {44}},\
  \bibinfo {pages} {L13} (\bibinfo {year} {1983})}\BibitemShut {NoStop}%
\bibitem [{\citenamefont {Nakayama}\ \emph {et~al.}(1994)\citenamefont
  {Nakayama}, \citenamefont {Yakubo},\ and\ \citenamefont
  {Orbach}}]{nakayama1994dynamical}%
  \BibitemOpen
  \bibfield  {author} {\bibinfo {author} {\bibfnamefont {T.}~\bibnamefont
  {Nakayama}}, \bibinfo {author} {\bibfnamefont {K.}~\bibnamefont {Yakubo}},\
  and\ \bibinfo {author} {\bibfnamefont {R.~L.}\ \bibnamefont {Orbach}},\
  }\bibfield  {title} {\bibinfo {title} {Dynamical properties of fractal
  networks: Scaling, numerical simulations, and physical realizations},\ }\href
  {https://doi.org/10.1103/RevModPhys.66.381} {\bibfield  {journal} {\bibinfo
  {journal} {Rev. Mod. Phys.}\ }\textbf {\bibinfo {volume} {66}},\ \bibinfo
  {pages} {381} (\bibinfo {year} {1994})}\BibitemShut {NoStop}%
\bibitem [{\citenamefont {Sapoval}\ \emph {et~al.}(1991)\citenamefont
  {Sapoval}, \citenamefont {Gobron},\ and\ \citenamefont
  {Margolina}}]{sapoval1991vibrations}%
  \BibitemOpen
  \bibfield  {author} {\bibinfo {author} {\bibfnamefont {B.}~\bibnamefont
  {Sapoval}}, \bibinfo {author} {\bibfnamefont {T.}~\bibnamefont {Gobron}},\
  and\ \bibinfo {author} {\bibfnamefont {A.}~\bibnamefont {Margolina}},\
  }\bibfield  {title} {\bibinfo {title} {Vibrations of fractal drums},\ }\href
  {https://doi.org/10.1103/PhysRevLett.67.2974} {\bibfield  {journal} {\bibinfo
   {journal} {Phys. Rev. Lett.}\ }\textbf {\bibinfo {volume} {67}},\ \bibinfo
  {pages} {2974} (\bibinfo {year} {1991})}\BibitemShut {NoStop}%
\end{thebibliography}
\end{document}